\documentclass[12pt,letterpaper]{JHEP3}

\input{epsf}
\usepackage{epsfig}
\usepackage{amsmath}
\usepackage{cite}

\def\pa{\partial}

\def\to{\rightarrow}
\def\be{\begin{equation}}
\def\ee{\end{equation}}
\def\bea{\begin{eqnarray}}
\def\eea{\end{eqnarray}}
\def\nonu{\nonumber \\{}}
\def\half{{1 \over 2}}
\def\cf{{\cal{F}}}

\def\cl{{\cal{L}}}
\def\cm{{\cal{M}}}
\def\ut{{\underline{t}}}
\def\ux{{\underline{x}}}
\def\ur{{\underline{r}}}
\def\uu{{\underline{u}}}
\def\utheta{{\underline{\theta}}}
\def\uphi{{\underline{\phi}}}
\def\upsi{{\underline{\psi}}}
\def\hut{{\hat{\underline{t}}}}
\def\hux{{\hat{\underline{x}}}}
\def\hutheta{{\hat{\underline{\theta}}}}
\def\huphi{{\hat{\underline{\phi}}}}

\def\sF{{{\rm F}\!\!\!\!\hskip.8pt\hbox{\raise1pt\hbox{/}}\,}}

\def\a{\alpha}

\def\d{\delta}
\def\e{\epsilon}

\def\f{\phi}

\def\k{\kappa}
\def\l{\lambda}
\def\m{\mu}
\def\n{\nu}

\def\p{\pi}

\def\r{\rho}
\def\s{\sigma}

\def\D{\Delta}
\def\F{\Phi}
\def\G{\Gamma}

\def\O{\Omega}
\def\P{\Pi}

\title{\begin{center}Open String Attractors\end{center}}
\author{\centerline{Joris Raeymaekers}\\
\centerline{Department of Physics, University of Tokyo,}
\centerline{Hongo 7-3-1, Bunkyo-ku, Tokyo 113-0033, Japan.}

\centerline{}
\centerline{}
\bigskip
\centerline{{\rm E-mail}:\email{joris@hep-th.phys.s.u-tokyo.ac.jp}}}

\abstract{We present a simple example of a supersymmetric attractor mechanism in the purely
open string context  of D-branes embedded in curved space-time. Our example involves a class of
D3-branes embedded in the 2-charge D1-D5 background of type IIB whose worldvolume contains a 2-sphere.
Turning on worldvolume fluxes, these branes carry induced $(p,q)$ string charges. Supersymmetric configurations
display a flow of the open string moduli towards an attractor solution  independent of their asymptotics.
The equations governing this mechanism closely resemble the  attractor
flow equations for supersymmetric
black holes in closed string theory. The BPS equations take the form of a gradient flow
and describe worldvolume solitons interpolating between an $AdS_2$ geometry where the two-sphere has collapsed,
and an attractor solution with $AdS_2 \times S^2$ geometry. In these limiting solutions, the preserved supersymmetry is enhanced from 4 to
8 supercharges.
We also discuss  the interpretation of our solutions as intersecting brane configurations placed in the D1-D5
background, as well as the S-duality transformation to the F1-NS5 background.}

\preprint{ hep-th/0702142 \\UT-07-07}

\begin{document}

\section{Introduction}

Extremal black holes in string theory have the property
that scalar moduli fields are drawn to fixed values at the horizon, which are determined
by the charges carried by the black hole.
This property, which  goes under the name of the attractor mechanism, has played an important role in the understanding
of black holes in string theory.  It  was first discovered in the context of
supersymmetric black holes in $N=2$ theories \cite{Ferrara:1995ih,Strominger:1996kf,Ferrara:1996dd} and more recently, has played a
crucial role in the formulation of the OSV-conjecture \cite{Ooguri:2004zv} and has been shown to apply to nonsupersymmetric
black holes as well \cite{Goldstein:2005hq,Tripathy:2005qp}. The physics underlying the mechanism is closely tied to the microscopic entropy
carried by the black hole: since the size of the horizon depends on the moduli, the latter
should approach values determined by the black hole charges and independent of their continuous asymptotic values.

General open-closed string duality considerations lead one to expect that
an attractor mechanism should exist for open string moduli as well. In the low-energy limit, open string dynamics
is described by D-brane effective actions consisting of Born-Infeld and Wess-Zumino terms, and closer inspection
shows that an attractor mechanism could occur in situations where both background and worldvolume gauge fields
are turned on. For example, consider a background containing $p$-branes producing a RR electric potential $C$ and
a $p+2$ brane probe wrapping a transverse
2-cycle with worldvolume magnetic field $F$ on this cycle. $F$ is quantized and represents a lower D-brane charge.
The term $\int F\wedge C $ in the worldvolume action
then represents a potential term for the scalar moduli that describe the D-brane embedding,
which is determined by the background and worldvolume charges and which vanishes
far away from the branes in the background.
It is therefore reasonable to expect that this interactions fixes a combination of
open string moduli  in terms of the worldvolume and background charges in the vicinity of the branes in the background.
This example can then be dualized to more general situations.
Such a mechanism is of course closely related to the stabilization of open string moduli in situations
with background and worldvolume fluxes, which was explored in \cite{Gomis:2005wc}, and ultimately goes back to
the observation of flux stabilization of D-branes \cite{Bachas:2000ik}.
We should stress that, although the appearance of an open string attractor mechanism seems plausible from the
explicit form of the interactions, it is not a priori clear whether there is an underlying
explanation in terms of an entropy contained in the open degrees of freedom. The open string attractor
mechanism is also expected to play an
important role in the open string version of the OSV conjecture proposed in \cite{Aganagic:2005dh}, which could  shed light
on a possible entropic interpretation.

In this work, we will describe in detail an explicit example of such an open string attractor mechanism.
We will consider here only the supersymmetric version,
although the above considerations suggest that, just like in the closed string case, the mechanism
is not restricted to the supersymmetric context.
Our main example will display a similarity to the supersymmetric attractor mechanism in the closed string context
that we find rather striking and deserving of a better explanation than we will be able to give at present.
Our example involves a class of D3-brane probes in the `D1-D5 system' (for a review, see \cite{David:2002wn}):
type IIB compactified on a fourfold $\cm$, with
D5-branes wrapping $\cm$ and coinciding with D1-branes in a noncompact direction, forming a six-dimensional black string.
 We consider D3-brane probes  where the worldvolume geometry contains a two sphere whose radius is allowed to
 vary over a 1+1 dimensional base.
 Such `spherically symmetric' configurations preserve an $SU(2)$ subgroup of the
target space isometry group. Turning on worldvolume electric and magnetic fluxes along the base and the $S^2$ fibre
respectively, the configurations carry fundamental and D-string charges and can be seen as
`fuzzy' $(p,q)$ string expanded to form a D3 brane through a form of the Myers effect. We derive
a BPS bound on the energy for D-brane configurations of this type, and show that the BPS-equations take
the form of a gradient flow.

In the near-horizon limit, the background geometry becomes $AdS_3 \times S^3$, and the BPS flow equations take
a form that is
remarkably similar to the attractor flow equations for supersymmetric black holes. Fixed points
of the flows correspond to extrema of a real and positive function
 $Z$. General flows represent worldvolume solitons which interpolate between a repulsive fixed point (a maximum of $Z$)
 at radial infinity,  corresponding to an $AdS_2$ worldvolume geometry where the $S^2$ has collapsed to zero size, and
the attractive fixed point (the minimum of $Z$) with $AdS_2 \times S^2$ worldvolume geometry near $r=0$.
The generic solution   preserves 4 of the `Poincar\'{e}' supersymmetries which extend to the full
asymptotically flat geometry, while for the fixed point solutions the supersymmetry is enhanced to 8 supercharges.
 The solution at the attractive fixed point can also be obtained by extremizing
an effective potential or `entropy' function \cite{Sen:2005wa}, whose physical meaning is less clear in this setting.
In the open string metric, the radii of the $AdS_2$ and $S^2$ factors become equal.

We will also investigate how our solutions extend to the full asymptotically flat background. This clarifies
their interpretation as intersecting brane configurations  placed
in the D1-D5 background. The general solution corresponds to a D3-brane transverse to the D1-D5 string in the background,
with a $(p,q)$ string `spike' running between the two. The transverse distance between the D3-brane and the D1-D5
string becomes the  asymptotic value of a modulus in the near-horizon limit. The near horizon solutions with
enhanced supersymmetry and $AdS_2$ or $AdS_2 \times S^2$ geometry correspond to the limiting cases where the
transverse distance is taken to infinity or zero respectively. We also discuss how our solutions transform
under S-duality to the F1-NS5 background.

Let us also comment on related D-brane solutions that have appeared in the literature.
The $AdS_2 \times S^2$ solution in the D1-D5 system was studied in \cite{Pawelczyk:2000hy,Couchoud:2003jw,Raeymaekers:2006np}. The S-dual
solution in the F1-NS5 background was first introduced in \cite{Bachas} and
has been studied extensively in the literature, as sampling of which is \cite{Bachas:2002nz}.
Similar D-brane solutions also exist
in the  Klebanov-Strassler \cite{Klebanov:2000hb}
and Maldacena-Nunez \cite{Maldacena:2000yy} backgrounds in the form of a $(p,q)$ string
expanding to form a  D3-brane wrapping an $S^2$ within the $S^3$ \cite{Herzog:2001fq}.
Our  solutions for general flows are related to the `baryon vertex' solutions and their generalizations
\cite{Witten:1998xy,Imamura:1998gk,Craps:1999nc,Pelc:2000kb,Camino:2001at}.

This paper is organized as follows. In section \ref{bpsder}, we introduce our brane configurations
and derive the BPS equations in both the asymptotically flat and near-horizon backgrounds. In section \ref{studynear},
we study the attractor flow equations in the near-horizon geometry and point out several analogies
with supersymmetric black hole attractors. Section \ref{asflatspace} discusses the extension of the
brane solutions to the asymptotically
flat spacetime and clarifies their interpretation. In section \ref{susyanal} we discuss the supersymmetries
preserved by our solutions, which provides an alternative derivation of the BPS equations. We
obtain the S-dual brane solutions in the F1-NS5 background in section \ref{sdualtransf} and end with a discussion
in section \ref{disc}. Appendix \ref{myers} clarifies the interpretation of our brane configurations as
$(p,q)$ strings expanded to a fuzzy D3-brane through a form of the Myers effect, while appendix \ref{killing}
gives a derivation of the Killing spinors of the background in our conventions.

\section{Spherical D3-branes in the D1-D5 background}\label{bpsder}
In this section we will set the stage for what is to be our main example of an open
string attractor. We will consider a class of BPS D3-branes  in the D1-D5 background, whose worldvolume geometry
has an $S^2$-fibre,
and derive the equations for their embedding into the background geometry from a bound on the energy.
Of course, the resulting system can also be derived from supersymmetry preservation, which we will do  in section \ref{susyanal}.
We will treat the full asymptotically flat background and the near-horizon limit simultaneously in this section,
providing a more detailed discussion for each case in later sections.
\subsection{Background}
We start by displaying our conventions for the D1-D5 background geometry. We consider type IIB on $\cm$ ($\cm$ being either
$K_3$ or $T^4$), with $Q_5$ D5-branes wrapped on $\cm$ and $Q_1$ D1-branes,
running parallel along a noncompact direction $x$. We choose spherical polar coordinates for the remaining transverse noncompact
directions. The string metric, dilaton and and RR three-form field strength are
\bea
ds^2 &=& (H_1 H_5)^{-1/2} ( - dt^2 + dx^2 ) + (H_1 H_5)^{1/2} (dr^2 + r^2 d\O_3^2) + \left( { H_1 \over H_5 } \right)^{1/2}
ds^2_\cm \nonu
e^{-\F} &=& {1 \over g}\left( { H_5 \over H_1 } \right)^{1/2}\nonu
F^{(3)} &=& {2 r_1^2 \over g r^3 H_1^2 } dt \wedge dx \wedge dr + {2 r_5^2 \over g}
 \sin^2 \psi  \sin \theta d\psi \wedge   d\theta \wedge d\f \label{background}
\eea
where $ds^2_\cm$ is the Ricci-flat metric on $\cm$ and
 $d\O_3^2$ is the metric on a unit $S^3$. We choose angular coordinates $\psi, \theta, \f$ on the $S^3$:
\be
d\O_3^2 = d \psi^2 + \sin^2 \psi ( d \theta^2 + \sin^2 \theta d\f^2 ).\label{metricS3}
\ee
where $\psi, \theta \in [0,\p],\ \f \in [0, 2 \p]$.
Note that the surfaces of constant $\psi$ are 2-spheres of radius $\sin \psi$. The harmonic functions
appearing in (\ref{background}) are
$$
H_{1,5} = a + {r_{1,5}^2 \over r^2}; \qquad r_1 = {4 \p^2 \a ' \over \sqrt{V_M}}  \sqrt{g Q_1 \a '},\qquad r_5 = \sqrt{g Q_5 \a '}
$$
where $a=1$ describes the asymptotically flat geometry while  taking $a=0$ gives the near-horizon
$AdS_3 \times S^3 \times \cm$ geometry in the Poincar\'{e} patch.
We will work in the following gauge for the RR two-form:
\be
C^{(2)} = {1 \over g H_1 } dt \wedge dx + { r_5^2 \over g}
( \psi - \sin \psi \cos \psi ) \sin \theta d\theta \wedge d\f
\ee
Note that there is a `Dirac string' singularity at $\psi=\p$, the invisibility of which imposes quantization
of $Q_5$.
The isometry group of the background is
\bea
ISO(1,1) \times SO(4) \qquad a=1\nonu
SO(2,2) \times SO(4) \qquad a=0\nonumber
\eea
Writing the $SO(4)$ as $SU(2) \times SU(2)$, it is the diagonal $SU(2)$ that acts transitively on the two-spheres
of constant $\psi$ in (\ref{metricS3}). This subgroup will play an important role in what follows.

\subsection{Spherical D3-branes}
We will now discuss a class of spherically symmetric D3-brane probes, carrying worldvolume flux,
in this background. We restrict our attention to branes whose worldvolume includes an $S^2$
embedded within $S^3$, parametrized by $\theta$ and $\f$ in (\ref{metricS3}), whose size we allow to
vary as a function of
the other coordinates. In other words, the worldvolume geometry is an $S^2$ fibered over a 1+1 dimensional base.
Such branes are `spherically symmetric' in the sense that
they preserve the  $SU(2)$ symmetry that acts on the $S^2$ fibre.
Such configurations naturally generalize the known D3-brane solutions in the near-horizon region with
$AdS_2 \times S^2$ geometry,
where the size of the $S^2$ is constant \cite{Pawelczyk:2000hy,Couchoud:2003jw,Raeymaekers:2006np}. The latter solutions will play a special role in what follows, as they will
play the role of the attractor geometry with enhanced supersymmetry.
We also allow general worldvolume gauge fields consistent with the $SU(2)$ symmetry. This restricts the
worldvolume gauge field to have an electric part $F_{\rm el}$ on the 1+1 dimensional base,
 a magnetic part $F_{\rm magn} $
with legs on the $S^2$ fiber, and no mixed components.

The terms contributing to the worldvolume action are then
\be S = - \m_3 \int d^4 \s e^{-\F} \sqrt{ - \det (\hat G + F ) } + \m_3 \int F_{\rm el} \wedge \hat C^{(2)}_{\rm magn}
+ \m_3 \int F_{\rm magn} \wedge \hat C^{(2)}_{\rm el}\label{abstractaction}\ee
where $\m_3 = 1/( (2\p)^3 \a '^2)$ is the D3-brane charge density and a $\hat \ $ denotes a pullback to the
worldvolume.
Turning on $F_{\rm el}$ is necessary for stabilizing the contractible $S^2$ on which the brane is wrapped.
With both $F_{\rm el}$ and $F_{\rm magn}$ turned on, the brane becomes a source for fundamental string charge
(denoted by $q$) and
D-string charge (denoted by $p$) as well. Let us first discuss the quantization conditions following from this.
Requiring that the source terms for the electric NSNS and RR two-forms are properly quantized leads to\footnote{
We are ignoring here curvature corrections to this formula, not to mention further subtleties in defining charges
in Ramond backgrounds. What we  find is, however, consistent upon S-dualizing with the better understood quantization
conditions in pure Neveu-Schwarz backgrounds, as we will see in section \ref{sdualtransf}.}
\bea
q  &=& {\m_3 \over \m_1} \int_{S^2} ( \star \tilde F_{\rm el} + \hat C^{(2)}_{\rm magn} )\label{f1quant}\\
p  &=& {\m_3 \over \m_1} \int_{S^2} F_{\rm magn}.\label{d1quant}
\eea
where $\star$ is the worldvolume Hodge star.
We have defined a field $\tilde F$ as
$$ \m_3 \sqrt{ - \det \;\hat G  }  \tilde F^{ab} = {\d S_{\rm BI} \over \d F_{ab}}$$
The integrals are to be performed over the $S^2$ fibre. The equation of motion and Bianchi
identity for the worldvolume gauge field imply that the charges are well-defined and  independent of the position on the base.
The fact that the fundamental string charge $q$ receives a Wess-Zumino contribution  from the second term in (\ref{f1quant})
has an important consequence in the near-horizon limit, where the $S^3$ becomes noncontractible, namely
that  $q$ takes values
in ${\bf Z}_{Q_5}$. The Wess-Zumino term is invariant under small gauge
transformations of $C^{(2)}$, but shifts by a multiple of $Q_5$ under large gauge transformations.
This is most easily seen
by writing it as ${\m_3 \over \m_1} \int_B \hat F^{(3)}$ with $B$ a 3-surface   chosen such that $\d B = S^2$.
Different choices of $B$ can
differ by a map with nonzero winding number $n$ around $S^3$ which, using the normalization of $F^{(3)}$ in (\ref{background}),
leads to an identification
\be q \sim q + n Q_5.\label{qperiod}\ee
As we shall illustrate in more detail in section \ref{sdualtransf},
this quantization condition is simply the S-dual version of the well-known D1-charge quantization in the background of
$NS5$-branes.  As long as the size of the $S^2$ fiber is sufficiently small,
we expect our configurations to describe $(p,q)$ strings expanded to form a `fuzzy' D3-brane through a form of the
Myers effect \cite{Myers:1999ps}.
We will come back to this fuzzy sphere description in more detail
shortly.

\subsection{Action}
We will now write out the action (\ref{abstractaction}) in more detail, starting by fixing the
worldvolume reparametrization invariance.
It will be convenient  to choose the worldvolume coordinates $\s^a$ to coincide with $(t, x, \theta, \f)$.
$SU(2)$ invariance restricts
the worldvolume scalars $ r, \psi$ and the electric field strength $F_{tx }$ to be independent
of $\theta, \f$, while $F_{\theta \f}$ should be proportional to $\sin \theta$.
The quantization condition (\ref{d1quant}) leads to
\be F_{\theta \f} = {p \m_1 \over 4 \p \m_3} \sin \theta \label{magnfield}\ee
where $\m_1 = 1/(2 \p \a ')$ is the D1-brane charge density. Substituting into the
 action and performing the $\theta , \f $ integrals leads to a consistent truncation of the original theory,
 resulting in an effective 1+1 dimensional action for a string-like object:
\be
S = -\m_1  \int dt dx \left[ p e^{-\F}
\sqrt{g \tilde g } - {Q_5 \over \p} F_{t x} (\psi - \sin \psi \cos \psi) -  {p  \over g   H_1}\right] \label{action}
\ee
where
\bea
g &\equiv& (H_1 H_5)^{-1} \left( 1  - H^1H_5 (\dot{r}^2 + r^2 \dot{\psi}^2 ) \right)
\left( 1  + H_1 H_5 ({r'}^2 + r^2 {\psi ' }^2 ) \right) - F_{t x}^2 \nonu
\tilde g &\equiv& 1 +{ r^4 H_1 H_5 \sin^4 \psi \over p^2 \p^2 \a'^2 }.\nonumber
\eea
Here, we denoted the time derivative by a $\dot{\ }$ and the $x$ derivative by a $\  '$.

We expect such configurations to represent
 $(p,q)$ strings expanded to form a `fuzzy' D3-brane through a form of the Myers effect \cite{Myers:1999ps}.
In an alternative description, they should arise as noncommutative fuzzy sphere solutions in the worldvolume theory of $p$
coinciding D-strings, with the $U(1)$  part of the worldvolume field strength turned to induce the fundamental string charge $q$.
From the analysis of \cite{Myers:1999ps} one expects the latter description (as a perturbative expansion in the matrix-valued coordinates)
to be valid as long as the $S^2$ radius in string units
is smaller than $p$, or
\be {\sqrt{H_1 H_5} r^2 \sin^2 \psi \over \p \a' } \ll p \label{overlapcond}. \ee
In this limit, we can expand $\sqrt {\tilde g}$ in (\ref{action}):
$$ \sqrt{ \tilde g } = 1 + { r^4 H_1 H_5 \sin^4 \psi \over 2p^2 \p^2 \a'^2 } + \ldots$$
In appendix \ref{myers} we show that  the noncommutative  worldvolume theory of $p$
coinciding D-strings allows a fuzzy sphere solution and, expanding around it, we  obtain precisely the action
(\ref{action})  in the limit (\ref{overlapcond}).

\subsection{Hamiltonian}
The canonical momenta $ P_r \equiv {\pa \cl \over \pa
\dot r },P_\psi \equiv {\pa \cl \over \pa
\dot \psi }$ and $\P \equiv {\pa \cl \over \pa
\dot F_{t x} }$
following from the action (\ref{action}) are given by:
\bea
P_r &=& \m_1 p  e^{-\F}
\sqrt{\tilde g \over g }\left( 1  + H_1 H_5 ({r'}^2 + r^2 {\psi ' }^2 ) \right)  \dot{r}\nonu
P_\psi &=& \m_1  p  e^{-\F}
\sqrt{\tilde g \over g }\left( 1  + H_1 H_5 ({r'}^2 + r^2 {\psi ' }^2 ) \right)r^2  \dot{\psi}\nonu
\P &=& \m_1  \left(  p  e^{-\F}
\sqrt{\tilde g \over g } F_{t x} + {Q_5 \over \p} (\psi - \sin \psi \cos \psi )\right)
\eea
We can define the phase-space lagrangian density $\cl$ as
\be \cl = \dot x P_x + \dot r P_r + \dot \psi P_\psi + \dot A_\r \P - L = \left(  \dot x P_x + \dot r P_r + \dot \psi P_\psi
+ F_{t\r} \P - L \right) - A_t \P' \label{hamdef}\ee
where we have done a partial integration in the second equality. The quantity in brackets can be identified as
the (improved) Hamiltonian density, while the second term imposes the Gauss law constraint $\P ' = 0$ for
the worldvolume gauge field.
The quantization condition (\ref{f1quant}) on the fundamental string charge fixes the
the integration constant in this equation
in terms of $q$:
$$\P = q \m_1.$$
Substituting in (\ref{hamdef}) and restricting attention to static configurations
$$ P_r = P_\psi = 0$$
one finds for the Hamiltonian
\be
H =  \m_1 \int d x \left[{Q_5 \over \p} \sqrt{ \D_1^2 + \D_2^2 + \D_3^2 } \sqrt{ {1 \over H_1 H_5 } + r'^2 + r^2
\psi'^2}
- {p  \over g  H_1 } \right].\label{energy}
\ee
Here, we have defined functions $\D_1, \D_2, \D_3$ that will reappear frequently in what follows:
\bea
\D_1 &\equiv &   \sin \psi \cos \psi - ( \psi - {q \over Q_5} \p )\nonu
\D_2 &\equiv &   {r^2 \over r_5^2} H_5 \sin^2 \psi \nonu
\D_3 &\equiv &    p {\p \over g Q_5} \sqrt{H_5 \over H_1 }\label{defdeltas}.
\eea
We also record, for later use, the expression for the worldvolume electric field
\be
F_{tx}= \sqrt{ {1 \over H_1 H_5 } + r'^2 + r^2 \psi'^2}{\D_1 \over \sqrt{\D_1^2 + \D_2^2+\D_3^2} }\label{elfield}.\ee
The total energy (\ref{energy}) is the sum of two competing contributions: the first term represents
the energy of a static string with variable tension
\be T(r,\psi) = \m_1 {Q_5 \over \p}  \sqrt{ \D_1^2 + \D_2^2 + \D_3^2 } .\label{tension}\ee
The second term represents the Coulomb energy of $p$ D-strings placed in the electric RR potential
$C^{(2)}_{\rm el}$  produced by the D-strings in the background. Both terms cancel
precisely for pure D1-string probes ($q=0,\ \psi=0$) placed parallel to
 the D1-strings in the
background (i.e. at constant $r$). This solution can be seen as the zero-energy ground state of the effective string
description.
Turning on the fundamental string charge $q$ amounts to turning on a central charge in
the worldvolume superalgebra, as we shall presently see, and leads to a class of BPS-solutions
that can be seen as worldvolume solitons.
When the D1-charge
$p$ is zero, the energy doesn't contain the Coulomb contribution and a BPS bound on the energy can be
derived using standard methods \cite{Camino:2001at}. We will comment on this
special case later on. When $p \neq 0$, the energy is not
manifestly a sum of positive contributions, but we will see that it is  possible to rewrite
it in an equivalent form suitable for deriving a BPS-type bound.

\subsection{BPS equations and gradient flow}
The Hamiltonian leads to  second order equations for $ r(x ), \psi(x )$, which will reduce
to a first order system for BPS configurations.
Some intuition can be gained by viewing the Hamiltonian  as an action functional describing geodesic motion
of  an effective particle with a position dependent mass $m(r, \psi) $ in a curved three dimensional space with metric
$dl^2= {1 \over H_1} dx^2 + H_5( dr^2 + r^2 d\psi^2)$  in the presence of a gauge potential
$A= - { p \over g H_1} dx$:
%in terms of a  worldline coordinate $\r$ with position dependent mass
%$m(r, \psi) $ in a curved three dimensional space with metric
%$dl^2$  in the presence of a gauge potential $A$:
\be H = {Q_5 \over \p } \m_1 \int d \r \left[ m(r, \psi ) \sqrt{ {{\dot x}^2\over H_1 } + H_5 ( \dot r^2 + r^2 \dot \psi^2)}
 - { p \p\over g Q_5 H_1} \dot x \right] \label{particleham}\ee
with
$$
m(r, \psi)=  \sqrt{ \D_1^2 + \D_2^2 + \D_3^2 \over H_5}\nonu
$$
We have introduced an arbitrary parameter $\r$ on  the worldline of the effective particle,
and denoted the $\r$-derivative by a $\dot \ $, hopefully without causing confusion with the time derivative (all
configurations considered henceforth will be static). We recover the
earlier expression (\ref{energy}) after choosing $\r = x$.
We can write a classically equivalent system without the square root
by introducing
an auxiliary  einbein $e(\r ) $ on the  worldline:
$$ H = {Q_5 \over \p } \m_1 \int d \r \left[ {1 \over 2 e}  \left({x'^2 \over H_1} + H_5( r'^2 + r^2 \psi'^2)\right)
+ {e \over 2}
{ \D_1^2 + \D_2^2 + \D_3^2 \over H_5} -  { p \p \over g Q_5 H_1} x'\right]$$
Solving for  for the einbein  leads back  to (\ref{particleham}).
This expression can, up to a boundary term, be written as a sum as a sum of squares. This can be seen by using the
definitions (\ref{defdeltas}) and observing that
$$
\D_1^2 + \D_2^2  = \left(\pa_{ r} ( r Z)\right)^2 + (\pa_\psi Z )^2
$$
where we have defined a function $Z$ as
\be
Z \equiv \sin \psi - ( \psi - {q \over Q_5} \p )\cos \psi  + { a \over 3 } \left({r \over r_5}\right)^2 \sin^3 \psi
\ee
The Hamiltonian can then be written as
\bea
H &=& {Q_5 \over  \p } \m_1  \int d\r  \Big[ \half \left( \sqrt{H_5\over e} \dot{ r} \pm
\sqrt{e  \over  H_5}\pa_{ r} ( r Z) \right)^2
+  \half \left( \sqrt{H_5 \over e} r \dot \psi \pm  \sqrt{e \over  H_5 }\pa_\psi  Z)\right)^2 \nonu
&&+ { 1 \over 2 e H_1}  \left( \dot x -   {p \p  \over g Q_5} e  \right)^2
 \mp {d \over d \r} (r Z)\Big]  \label{asflatham}
\eea
The energy is bounded below by a total derivative, and configurations saturating the bound will automatically
obey the equations of motion. This gives the desired system of first order BPS equations.
The quantity ${Q_5 \over  \p } \m_1  r Z$ plays the role
of a central charge in the worldvolume superalgebra \cite{Bergshoeff:1997bh,Gauntlett:1997ss,Craps:1999nc}.

Let us first consider the case $p\neq 0$. It will be convenient to define dimensionless parameters
$\tilde x$, $ \tilde r$ as
$$ \tilde x \equiv {g Q_5 \over p \p  r_5} x; \qquad \tilde r \equiv {r \over r_5} .$$
Choosing $\r = \tilde x$ in (\ref{asflatham}), the BPS equations reduce to $e=r_5$ and
\bea
  \dot{ \tilde r} &=&  - {1 \over H_5} \pa_{\tilde r} ( \tilde r Z) \nonu
 \dot{\psi} &=& - {1 \over H_5 \tilde r^2} \pa_\psi ( \tilde r Z) \label{asflatbpseqs}.
\eea
Here we have chosen the upper sign in (\ref{asflatham}) without loss of generality, as
 the equations with the other sign choice are
related by a reflection $\tilde x \to -\tilde x$.
The equations describe a gradient flow with potential function $\tilde r Z$ on a space with
 metric $H_5 (d \tilde r^2 + \tilde r^2 d
\psi^2 ) $.
It will be useful at times to change the independent variable to $\tilde r$ and obtain equations for $\tilde x( \tilde r),
\psi (\tilde r)$:
\bea
{d \tilde x \over dr } &=& -  {H_5 \over \pa_{\tilde r} (\tilde r Z)} \nonu
\tilde r {d \psi \over d \tilde r } &=& {\pa_\psi Z \over \pa_{\tilde r} (\tilde r Z)}.\label{reqs}
\eea
The total energy of the solutions is
\be E =  {Q_5 r_5 \over  \p } \m_1  \left[ \tilde r Z\right]^{\tilde x_i}_{\tilde x_f} .\label{energyfull}  \ee

We now turn to the case $p=0$. In this case, we cannot choose $\rho$ to be proportional to
$x$ anymore. Instead,
we can take $\r = \tilde r$, and the BPS equations become
\bea
{d  x \over dr } &=& 0 \nonu
\tilde r {d \psi \over d \tilde r } &=& {\pa_\psi Z \over \pa_{\tilde r} (\tilde r Z)}.\label{pzerosol}
\eea
Comparing with (\ref{reqs}), we see that the flows in the $p=0$ case are simply the flows for any $p\neq 0$ projected
onto a surface of constant $x$.

\subsection{Worldvolume geometry}
We collect here for later convenience  also  the formulae for the worldvolume fields for  BPS solutions satisfying (\ref{asflatbpseqs}).
Making use of the relation $H_1 H_5 (r'^2 + r^2 \psi'^2 ) = {\D_1^2 + \D_2^2\over \D_3^2}$ satisfied by these solutions, the induced metric and the gauge field on the worldvolume can be written as
\bea
d \hat s^2 &=& (H_1 H_5)^{-1/2} \left( - dt ^2 + {  \D_1^2 + \D_2^2  +\D_3^2 \over \D_3^2} d x^2 \right)
+ (H_1 H_5)^{1/2} r^2 \sin^2 \psi \left( d \theta ^2 + \sin^2 \theta d \f^2\right)\nonu
F &=& (H_1 H_5)^{-1/2} {\D_1 \over \D_3} dt \wedge d x + (H_1 H_5)^{1/2} r^2 \sin^2 \psi  {\D_3 \over \D_2}
\sin \theta d \theta \wedge d\f \label{indmetric}
\eea
One easily derives the components of the open string metric $g^o_{\m\n} = g_{\m\n} - \cf_{\m\r} g^{\r\s}\cf_{\s\n}$:
\be
d \hat s_o^2 = {\D_2^2 + \D_3^2\over \sqrt{H_1 H_5}}  \left( - {dt^2 \over  \D_1^2 + \D_2^2  +\D_3^2}
+  { dx^2  \over \D_3^2}\right)
 + {\sqrt{H_1 H_5}(\D_2^2 + \D_3^2) r^2 \sin^2 \psi\over  \D_2^2} \left( d\theta^2 + \sin^2 \theta d\f^2 \right)
\label{openmetric}
 \ee

\section{The near-horizon limit and attractor flows}\label{studynear}

We shall now discuss the solutions of the BPS equations (\ref{asflatbpseqs}) in the near-horizon limit of the
background geometry obtained by putting $a=0$ in the equations above. It is in this limit that
the BPS flow equations most closely resemble the attractor flows for supersymmetric black holes in $N=2$ supergravity.

\subsection{Flow equations}
A first observation is that, in the near-horizon limit,  the function $Z$ depends on $\psi$ alone:
$$ Z = Z(\psi) = \sin \psi - ( \psi - {q \over Q_5} \p )\cos \psi . $$
As we saw in (\ref{qperiod}), $q$ is a ${\bf Z}_{Q_5}$ valued charge and we will  take $0 \leq q <  Q_5$ in what
follows.
The function $Z$  is then a positive function with  a single minimum at $\psi = {q \over Q_5 }\p \equiv \psi_* $ and
maxima at the boundary points $\psi = 0,\p$ (see
figure \ref{Z+pot}(a)).
\FIGURE{\begin{picture}(500,150)
\put(35,15){0}\put(25,110){$Z$}\put(105,15){$\psi_*$}\put(160,15){$\psi$}\put(190,15){$\p$}
\put(270,20){$\psi$}\put(370,30){$-e^{U}$}\put(395,110){$V$}
\put(100,0){(a)} \put(250,0){(b)}
\put(40,30){\epsfig{file=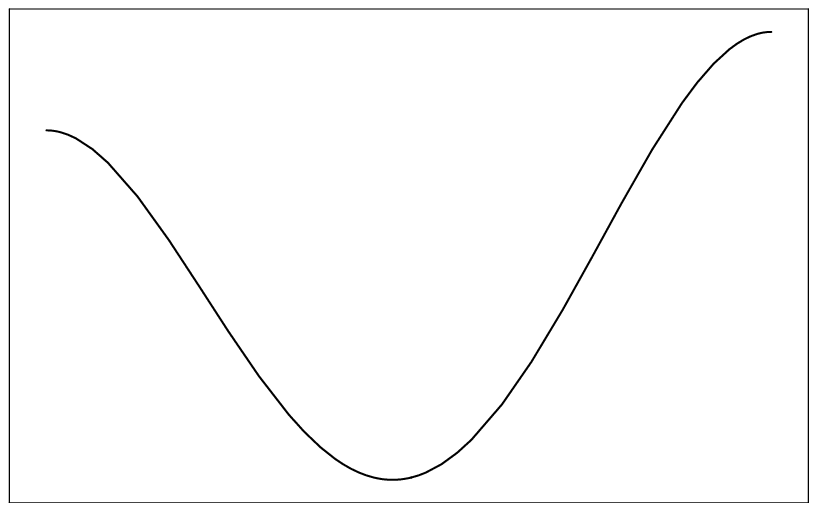, width=150pt }}
\put(240,0){\epsfig{file=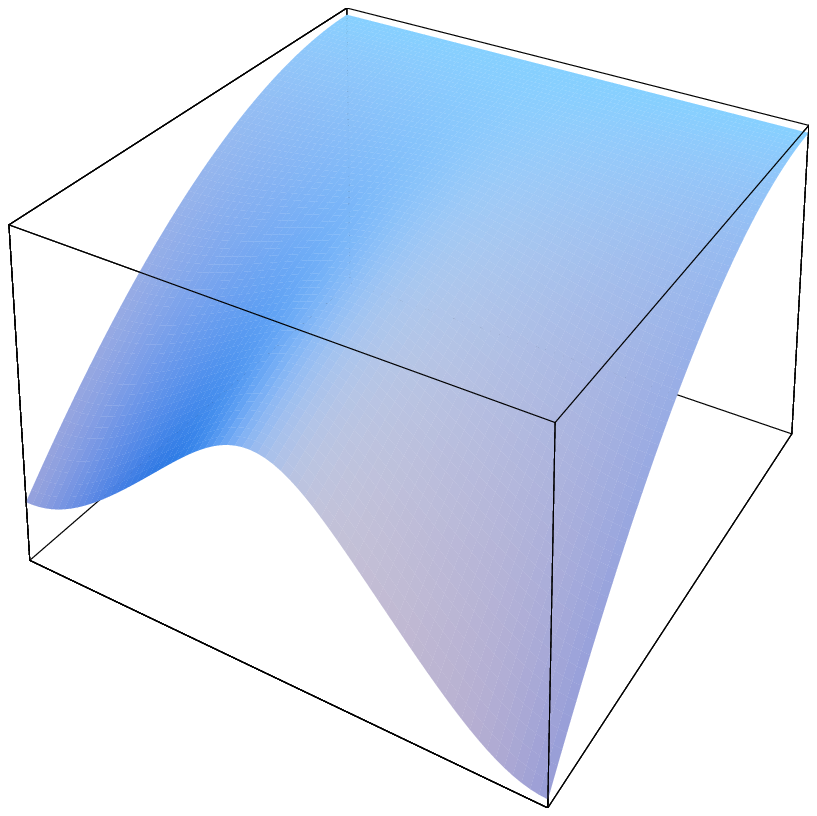, width=150pt }}
\end{picture}\caption{(a) The function $Z$. (b) The  potential $V$.}\label{Z+pot}}
It will be convenient to rewrite the equations derived in the previous section in terms of a coordinate $U$ defined as
$$ \tilde r \equiv e^U. $$
Assuming $p\neq 0$, it is a consistent truncation to substitute the value $ e = r_5$, which solves the equation of
motion for $\tilde x$, into (\ref{asflatham}) as long as we impose the equation for $e$ as a constraint.
Choosing again $\r = \tilde x$, this truncated energy function takes the form:
\bea
H &=& {\m_1 Q_5 r_5 \over 2 \p}  \int d\tilde x  \left[ \dot{U}^2 +  \dot \psi^2
+ e^{2U}  ( Z^2 + \pa_\psi Z^2 )\right]\nonu
&=& {\m_1 Q_5 r_5 \over 2 \p}  \int d\tilde x  \left[  \left(  \dot{U} \pm
 \pa_{ U} (e^U Z) \right)^2
+  \left( \dot \psi \pm \pa_\psi (e^U Z) \right)^2 \right]
\pm {\m_1 Q_5 r_5 \over \p } [e^U Z ]^{\tilde x_i}_{\tilde x_f}  \label{nearhorham}
\eea
It describes a particle moving on the $(U , \psi)$ strip with flat metric in an inverted  potential
$V = - e^{2U}  ( Z^2 + \pa_\psi Z^2 )$ (see figure \ref{Z+pot}(b)).
The constraint from the equation for $e$ becomes
\be  \dot{U}^2 +  \dot \psi^2
- e^{2U}  ( Z^2 + \pa_\psi Z^2 )  = 0 \label{nearhorconstraint} .\ee
It states that the conserved total `energy' of the effective particle is zero and can be imposed as an initial
condition.
Choosing again the upper sign, the BPS equations are
\bea
\dot U  &=&  - \pa_{U} (e^U  Z) \label{gradientflow1}\\
\dot{\psi} &=& - \pa_\psi(e^U  Z). \label{gradientflow2}
\eea
Note that solutions to these equations obey the constraint (\ref{nearhorconstraint}). The energy is given by
\be E =  {Q_5 r_5 \over  \p } \m_1  \left[ e^UZ\right]^{\tilde x_i}_{\tilde x_f} .\label{energynear}  \ee

\subsection{Solutions}
The system (\ref{gradientflow1}, \ref{gradientflow2}) describes a gradient flow on the flat $(U,\psi )$ strip with  potential function $e^U Z$.
The flow is directed towards the minimum of $Z$, since the second equation implies that
$$ \dot Z = - e^U (\pa_\psi Z)^2 \leq 0 $$
Hence there is an attractive fixed point at
$$ \psi_* = {q \over Q_5} \p$$ where $Z$ is minimal and takes the value $Z_* = \sin {q \over Q_5} \p $.
The maxima of $Z$ at $\psi = 0$ or $\p$ represent repulsive fixed points.
Furthermore, $U$ is a decreasing function since
$ \dot U = - e^U Z \leq 0 $ and the flows will eventually end up at $U_* = -\infty$, corresponding to
the `horizon' $r=0$ in Poincar\'{e} coordinates. The BPS solutions correspond to particle trajectories
where the initial conditions are tuned such that the particle reaches  the top of the inverted potential asymptotically.
Figure \ref{flows} illustrates these aspects of the gradient flow.
\begin{center}\FIGURE{\begin{picture}(500,150)
\put(10,20){0}\put(90,0){(a)}\put(160,20){-U}\put(10,90){$\psi$}\put(10,120){$\p$}\put(250,0){(b)}
\put(230,20){$\psi$}\put(330,30){$-e^{U}$}\put(355,110){$V$}
\put(20,30){\epsfig{file=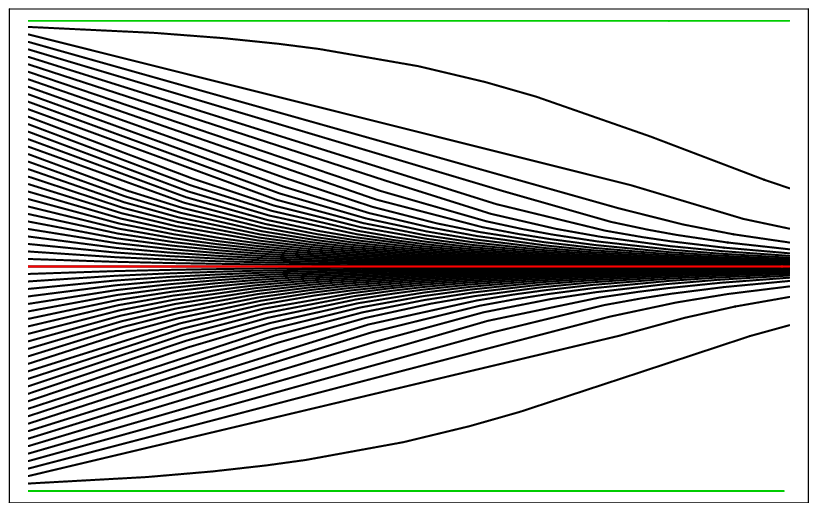, width=150pt }}
\put(200,0){\epsfig{file=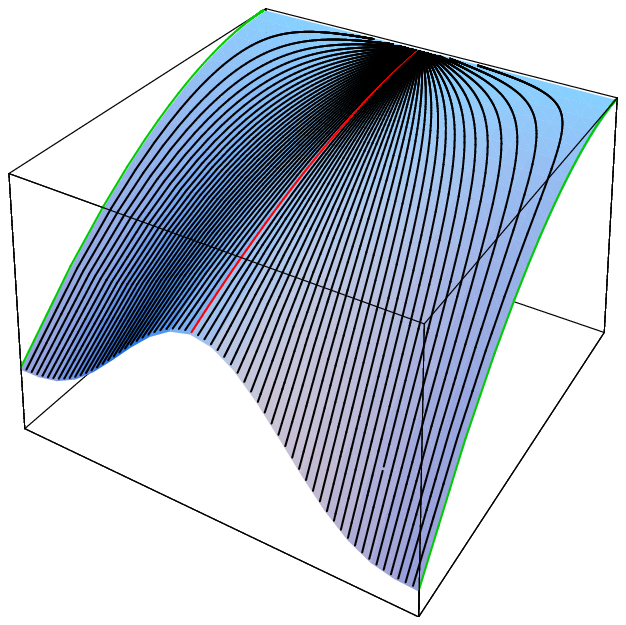, height=150pt }}
\end{picture}\caption{(a) Gradient flows in the $(-U,\psi)$ plane. The red line is the attractive fixed point,
the green lines are repulsive fixed points. (b) BPS trajectories `shoot for the top' of the inverted potential.}
\label{flows}}
\end{center}
In terms of the embedding of the fuzzy $(p,q)$ string, these equations tell us that, if we fix one endpoint of the
string somewhere in $AdS_3$ and specify some value of the fuzzy radius at this point, while letting the other end
`flap in the breeze',
the string will eventually reach $r=0$ at $ x = \infty$, and the $S^2$ radius will approach the value $\sqrt{r_1 r_5} Z_*$.
The equations (\ref{gradientflow1}, \ref{gradientflow2}) can also be solved exactly and we will now discuss
the different types of solutions in more detail.
We are  interested
in complete flows where the string starts out at the boundary of of $AdS_3$ and not somewhere  in the interior,
which would be forbidden by charge conservation.
We will use translation invariance in the $x$-direction to make the  starting point at $r= \infty$ correspond to $x=0$, which fixes
the integration constant in (\ref{gradientflow1}). We also note that  the equations (\ref{gradientflow1}, \ref{gradientflow2})
are invariant under changing $\psi \to \p - \psi$ and $q \to Q_5 -q$.

\subsubsection{Attractive fixed point: $AdS_2 \times S^2$ branes}
Of special importance is the solution where $\psi$ takes on the constant attractor value $\psi_*$ everywhere:
\bea
\psi &=& \psi_* \nonu
\tilde r &=&{1 \over \sin \psi_* \tilde x }\label{nearhorattrsol}
\eea
Generic solutions of  (\ref{gradientflow1}, \ref{gradientflow2}) approach this one for large $\tilde x$ and we
will call this the `attractor solution'.
The tension (\ref{tension}) for this solution is constant and given by
$$ T_{(p,q)}^{\rm attr} =  \m_1 \sqrt{ p^2 e^{-2 \f} + \left( {Q_5  \over \p } \sin {q  \p \over Q_5 }\right)^2}$$
The induced metric (\ref{indmetric}) for this solution is $AdS_2 \times S^2$, and we will see in section \ref{susyanal}
(see also \cite{Raeymaekers:2006np}) that it is $1/2$-BPS,
preserving 4 out of  8 Poincar\'{e} supercharges and 4 out of 8 conformal supercharges  of the near-horizon background.
In the induced metric on the worldvolume (\ref{indmetric}),
the radii of the $S^2$ and $AdS_2$ factors are different and given by
\be R_{\rm AdS_2} = L {T_{(p,q)}^{\rm attr} \over T_{(0,q)}^{\rm attr}} ;\qquad R_{\rm S^2}
= L Z_*\label{indradii}\ee
with $L\equiv \sqrt{r_1 r_5}$.
An interesting feature of the attractor solution is that, in the open string metric (\ref{openmetric}), the $S^2$ and $AdS_2$
radii become equal and are given by
\be
R_{\rm AdS_2}^o = R_{\rm S_2}^o = L  { \sqrt{ T_{(p,q)}^{{\rm attr\ }2} \sin^2 \psi_* +  T_{(p,0)}^{{\rm attr\ }2} \cos^2 \psi_*}
\over T_{(0,q)}^{\rm attr} }.
\ee
This property is consistent with an argument made in the S-dual system  in \cite{Bachas}.

\subsubsection{Repulsive fixed point: $AdS_2$ branes}
Another special solution to the equations (\ref{gradientflow1}, \ref{gradientflow2})  which deserves to be mentioned is
obtained by taking $\psi$ to  be constant and
equal to $0$ or $\pi$, the maxima of $Z$. These solutions correspond to the repulsive fixed points of the
flow equations, and small supersymmetric deformations cause the flow to move away from them and end up at
 the attractive fixed point. One could also wonder whether these solutions are physical, since
the $S^2$ fiber has collapsed to a point and it is not clear whether the D3-brane description is still
reliable.
Nevertheless, in the description as a noncommutative theory on coinciding D-strings,
 they simply correspond to the solution with commuting matrices discussed in appendix \ref{myers}, and that
 formulation should be
reliable. In section \ref{asflatspace}, we shall show that these solutions
arise as `spikes' on a D3-brane in the limit that the D3-brane is moved away to  infinity.
These observations suggest that we should not discard these solutions.
 The $\psi = 0$ solution has the  tension (\ref{tension})  of a $(p,q)$ string in
flat space
$$ T_{(p,q)} = \m_1 \sqrt{ p^2 e^{-2 \f} + q^2} $$
(for $\psi = \p$ we have to replace $q$ by $Q_5 - q$).
The radial coordinate is given by
\be \tilde r = {1 \over \psi_*} {1 \over \tilde x} \label{repsol}\ee
and the resulting worldvolume geometry is $AdS_2$, with the
radius in the induced and open string metrics given by
$$  R_{\rm AdS_2} = L {T_{(p,q)}\over T_{(0,q)}} \qquad
 R_{\rm AdS_2}^o = L {T_{(p,0)} \over T_{(0,q)}}$$
These solutions can be shown to preserve half of the near-horizon  supersymmetries as well \cite{Raeymaekers:2006np}.

These solutions represent a $(p,q)$ string
which has not expanded to form a D3-brane.
The situation is reminiscent of the case of giant gravitons \cite{McGreevy:2000cw,Grisaru:2000zn}, where there are also different supersymmetric
solutions representing expanded and non-expanded configurations carrying the same charges. This is
perhaps not so surprising, since the coupling which allows the two-sphere to be stabilized in our case
(i.e. the coupling of the electric
worldvolume field to the RR background) is T-dual to the coupling  allowing giant gravitons to expand (i.e. the coupling
of an  angular momentum to the RR background).

\subsubsection{General flows}
The general solution  to the flow equations (\ref{gradientflow1}, \ref{gradientflow2}) is
\bea
\psi &=& \psi_* + {1 \over C_1 \tilde x + C_2}\nonu
\tilde r &=& C_1 {\psi - \psi_* \over \sin \psi }\label{nearhorsol}
\eea
The solution breaks scale and translation symmetry of the background, and solutions with different values of the
 integration constants $C_1,\ C_2$ are related by the action of these broken generators.
The initial condition that  $r= \infty$ at $\tilde x=0$  fixes the constant $C_2$ in
(\ref{nearhorsol}).  We see from (\ref{nearhorsol}) that the flows must start
at either $\psi = 0$ or $\psi = \p$. Due to the above mentioned symmetry
$\psi \to \p - \psi$, $q \to Q_5 -q$ we can restrict attention to flows starting at $\psi = 0$.
These have $C_2 = - 1/\psi_*$, $C_1$ negative and $\psi \leq \psi_*$ everywhere.
The  constant
$C_1$ represents the  value of $r \sin \psi$ at the boundary of $AdS_3$ and can be interpreted as an asymptotic modulus.
As we will see in section \ref{susyanal}, the generic solution is 1/4 BPS, preserving 4 out of the 16 real supercharges
of the near-horizon region. It preserves half of the Poincar\'{e} supersymmetries but breaks all of the conformal ones.
A plot of various flows in  coordinate space is shown in figures \ref{exampleflow}(a) and \ref{coordflows}(a).
\FIGURE{\begin{picture}(300,200)
\put(60,0){(a)}\put(222,0){(b)}
\put(15,25){$\tilde r \cos \psi$}\put(175,25){$\tilde r \cos \psi$}
\put(120,40){$\tilde r \sin \psi$}\put(280,40){$\tilde r \sin \psi$}
\put(-10,150){$\tilde x$}\put(152,150){$\tilde x$}
\put(-10,10){\epsfig{file=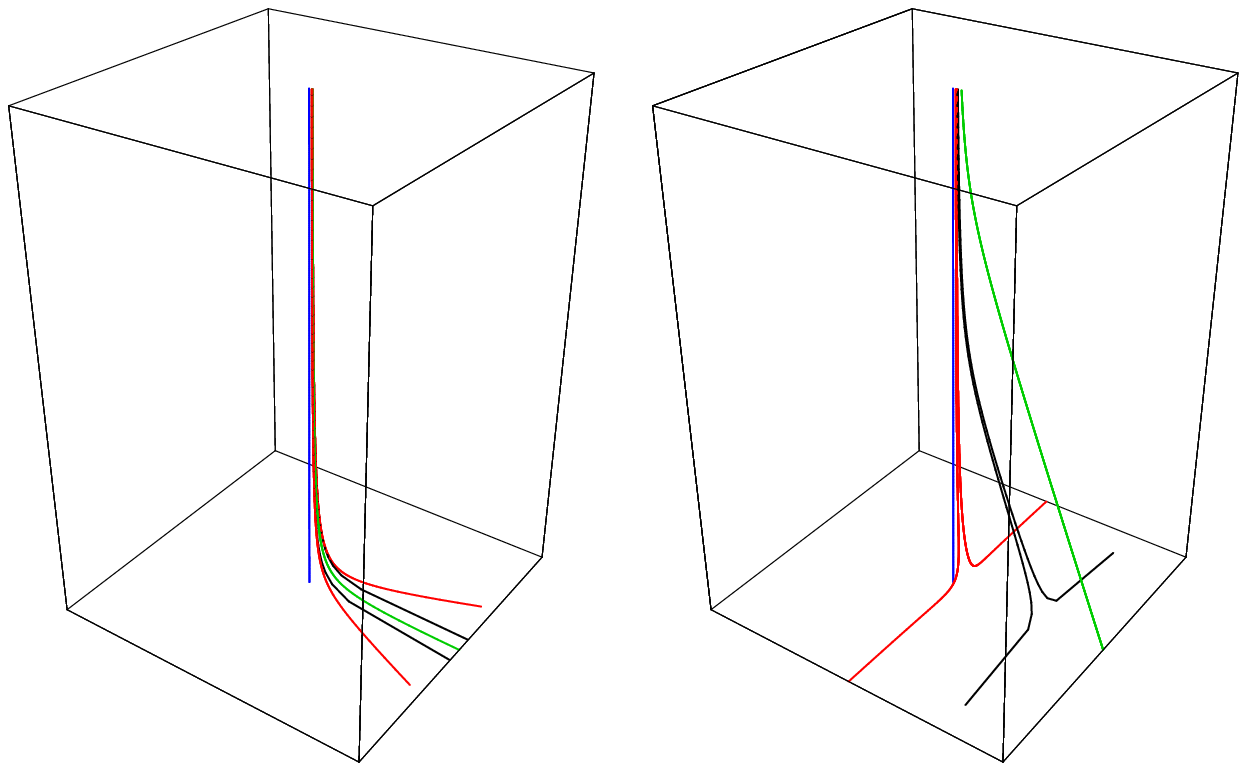, height=200pt }}
\end{picture}\caption{Flow plots in $(\tilde r \cos \psi, \tilde r \sin \psi, \tilde x )$ coordinates. Each
 brane configuration forms a `tube' whose cross-section is an $S^2$, represented  by two points on opposite sides of the
 $\tilde r \sin \psi$ axis. The vertical blue line represents the D1-D5 string in the background. (a) Solutions in the near-horizon geometry: the black curve denotes a generic flow, while the green and
 red curves represent the repulsive and attractive fixed point solutions with $\psi = 0$ and $\psi = \psi_*$ respectively.
 (b) Solutions in the full geometry: the generic solution represents a $(p,q)$ string `spike' ending on a D3-brane transverse
 to the D1-D5 string in the background.
 The attractive  and repulsive fixed point solutions arise from the limits where the transverse distance of the
 D3-brane is taken to zero or infinity respectively.
}\label{exampleflow}}

The attractor solution
 (\ref{nearhorattrsol}) and the repulsive solution (\ref{repsol}) are  obtained in the limits $C_1 \to - \infty$,
 $C_1 \to 0$ respectively, where scale invariance is restored. Hence we see that the general 1/4 BPS flows
 represent worldvolume solitons that
  interpolate between the
1/2 BPS repulsive solution at $\tilde x =0 $ and the 1/2 BPS attractor solution at $\tilde x = \infty$.
The tension (\ref{tension}) also interpolates between $T_{(p,q)}$ and $T_{(p,q)}^{\rm attr}$ as shown in
 figure \ref{tensionplot}.\\
\FIGURE{\begin{picture}(300,100)
\put(35,65){$T$}\put(260,-10){$x$}
\put(45,5){$T_{(p,q)}^{\rm attr}$}\put(45,115){$T_{(p,q)}$}
\put(70,0){\epsfig{file=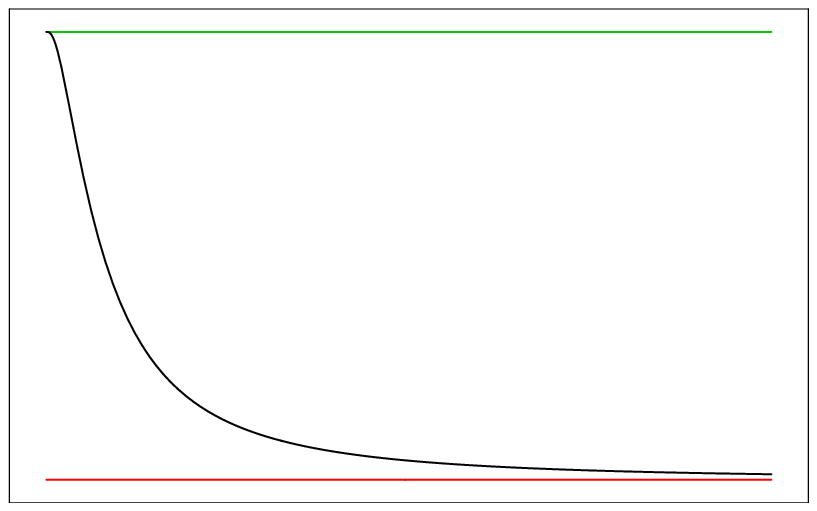, width=200pt }}
\end{picture}\caption{The tension of a generic flow solution (black line) interpolates between the tension
of the repulsive fixed point (green line) and attractive fixed point (red line) solutions.}\label{tensionplot}}

The energy of the solutions can be read off from (\ref{energynear}). Since
$e^U Z$ becomes zero for $\tilde x \to \infty$ it is given by
$$E  =  {\m_1 Q_5 r_5 \over \p } [e^U Z]_{|\tilde x =0}$$
The energy  is divergent due to the fact that the string stretches all the way to
the  boundary of $AdS_3$. The variable $\psi$ approaches zero near the boundary and introducing a cutoff at
a small value of  $\psi$ we find
\bea
E &=&  {\m_1 Q_5  \over \p } \lim_{\psi \to 0} r(\psi) \left( \sin \psi - (\psi  - \psi_*) \cos \psi  \right)\nonu
&=& {q \m_1} \lim_{\psi \to 0} r(\psi)
\eea
We see that the energy is equal to the energy  of $q$ fundamental strings stretched along the radial direction,
perpendicular to the D1-D5 string in the background. We note in particular that the regularized energy doesn't depend on the
asymptotic modulus $C_1$ and the solutions are degenerate in this sense. The D-string charge $p$ doesn't enter into
the expression for the energy because  a D-string probe is mutually BPS with the branes in the background.

Let us also comment on the solutions with zero D1-charge $p=0$, which, as we saw in (\ref{pzerosol}),
are obtained by projecting
the $p\neq 0$  solutions onto a surface of constant $x$, as illustrated in figure \ref{coordflows}(a). The  BPS equation (\ref{pzerosol}) is the same one that arises in
the description of D3-branes with electric field in a D5-brane background  and has been studied in the
literature before \cite{Camino:2001at}. It is  a  special case of a class of generalized
`baryon vertex' solutions that were studied in \cite{Witten:1998xy,Imamura:1998gk,Craps:1999nc,Pelc:2000kb}.
It was observed in these works that such solutions approach
a special solution where the angle $\psi$  is constant. We can now reinterpret this property as a special
(albeit less transparent) case of the attractor mechanism. It would  be interesting to study the
question of supersymmetry enhancement at the attractor point in these examples.

Our discussion of the solutions has been entirely in the Poincar\'{e} patch, and it would be interesting to study
how the geometry extends into global $AdS_3$ in more detail. While both the  attractive and repulsive
fixed point solutions represent static configurations with respect to global time, this no longer the case for
the general solution which will interpolate between the two in a time dependent manner.

\subsection{`Entropy' function}
For attractor black holes, the attractor geometry where the moduli take their constant fixed-point values
can also be derived by extremizing an entropy function \cite{Sen:2005wa}, which, at the minimum, coincides with
the physical entropy of the black hole. In our open string example, a similar role is played by the energy function
(\ref{asflatham}) evaluated for an ansatz where the worldvolume geometry is $AdS_2 \times S^2$. Extremizing this function
yields the value of the $AdS_2$ and $S^2$ radii in terms of the charges, as we shall presently illustrate. It's not
clear whether it can be related to an entropy contained in
the open string degrees of freedom, which is one of the main questions raised by our example.

We start by defining new target space coordinates $V_1, V_2$:
\bea
V_1 &=& \sqrt{ 1 + (ux)^2 }\nonu
V_2 &=& \sin \psi\nonumber
\eea
with $u \equiv {r \over L^2},\ L \equiv \sqrt{r_1 r_5}$.
If $(V_1, V_2)$ take on constant values $(v_1,  v_2)$, the induced worldvolume metric
is $AdS_2 \times S^2$:
$$ d \hat s ^2 = L^2 \left[ - u^2 dt^2 + {v_1^2 \over u^2} du ^2 + v_2^2 (d \theta^2 + \sin^2 \theta d\f^2)\right]. $$
Hence $v_1$ and $v_2$ are the radii of $AdS_2$ and $S^2$ in units of $L$. In analogy with the entropy
function for black holes, we define the entropy function $F(v_1, v_2; p,q)$ to be
the Hamiltonian density (\ref{asflatham}) evaluated at constant values of the scalar fields $V_1,\ V_2$:
$$
H \equiv \int d u F(v_1, v_2; p,q).
$$
One finds
$$
F(v_1, v_2; p,q) = L^2 \left[ T (v_2) v_1  - T_{(p,0)}^{\rm attr} \sqrt{v_1^2 - 1}\right]
$$
where $T$ is the tension given in (\ref{tension}). Extremizing the entropy function  gives
the correct values for $v_1,\ v_2$ at the attractor point. The variation with respect to $v_2$
tells us to extremize the tension and determines $$v_2 = \sin \psi_*.$$ Variation with respect to $v_1$
yields
$$ v_1 = {T_{(p,q)}^{\rm attr} \over T_{(0,q)}^{\rm attr}}$$
in agreement with the earlier result (\ref{indradii}). We note that the value of the entropy function at the minimum
is independent of the probe D1-charge and is given by
$$ F = L^2 T_{(0,q)}^{\rm attr} .$$
In the case of attractor black holes, the entropy function formalism greatly facilitates finding the attractor solution
in the presence of higher derivative corrections \cite{Sen:2005wa}, and it would be interesting to see if the same is
true here.

\subsection{Comparison with the attractor mechanism for  black holes }

The attractor mechanism described above is remarkably similar to the familiar attractor mechanism governing
 supersymmetric black holes in $N=2$ supergravity theories with vector multiplets \cite{Ferrara:1995ih,Strominger:1996kf,Ferrara:1996dd}.
Let us pause for a moment to identify similar quantities appearing in both systems.

Spherically symmetric attractor black holes in $N=2$ supergravity theories are described by an effective particle action
 and constraint of the form
(\ref{nearhorham},\ref{nearhorconstraint}) \cite{Ferrara:1997tw,Denef:2000nb}.
The  flow equations that describe the evolution of the spacetime metric and the moduli  take the form
of a gradient flow analogous to (\ref{gradientflow1}, \ref{gradientflow2}):
\bea
\dot U  &=&  - \pa_{U} (e^U  |Z|) \nonu
\dot{z^a} &=& - g^{a \bar b} \pa_{\bar b}(e^U  |Z|). \label{sugraflow}
\eea
Here, the function $U$ appears in the metric ansatz
$ds^2 = - e^{2U} dt^2 + e^{-2U} d{\bf x}^2$, the $z^a$ are complex vector multiplet moduli and $g_{a \bar b}$ is
the moduli space metric. The flow parameter is proportional to the inverse of the radial coordinate.
The function $Z$ is the graviphoton charge, which plays the role of the central charge in the $N=2$ superalgebra.
The equations describe a flow towards a minimum of $Z$, which becomes proportional to the black hole
horizon radius. General solutions
interpolate between the Minkowski vacuum at asymptotic infinity and
 an  attractor solution where the moduli take on constant values
and the geometry  is the $AdS_2 \times S^2$ Bertotti-Robinson solution near the horizon.
The general solution preserves 4  supersymmetries and interpolates between maximally supersymmetric vacua
that preserve all 8 supersymmetries.
The attractor geometry can also be derived from extremizing
an `entropy function' whose value at the extremum is the black hole entropy.

It's easy to draw parallels with our open string example. The variable $U$ is now related to
 the time component of the induced worldvolume metric: $ \hat g_{tt} \sim - e^{2U}$. The vector multiplet moduli are replaced
in our example by a single real field $\psi$, which is related to the size of the $S^2$ fiber. The role of the
 graviphoton charge is played by the real, positive function $Z$, which, at its minimum, is proportional
 to the size of the $S^2$ \footnote{One should note however that the function $Z$ is not quite the same as the
 worldvolume central charge, which is
 instead given by $e^U Z$.}. General flows preserve 4 supersymmetries and interpolate between solutions where
 the supersymmetry is enhanced to 8 supercharges: at infinity, an $AdS_2$ geometry where the $S^2$ has collapsed, and
near $r=0$, an $AdS_2 \times S^2$ geometry. The latter `attractor solution' can also be obtained by extremizing
an `entropy' function as we saw in the previous paragraph.

\section{Solutions in the asymptotically flat background}\label{asflatspace}
We now describe how the above solutions extend to the full, asymptotically flat
D1-D5 geometry.
This will help  clarify the interpretation of our brane solutions as intersecting D3-brane `spike' configurations
 embedded in the D1-D5 background. It will also provide a physical interpretation
of the asymptotic modulus $C_1$ encountered in the near-horizon solution (\ref{nearhorsol}).

The BPS equations are now given by (\ref{asflatbpseqs}) for $a=1$ and
 can still be solved analytically.
For this, it's convenient to switch to $\psi$ as the independent variable and solve for $\tilde r (\psi),
\tilde x (\psi ) $. The general solution satisfies
\bea
{\tilde r \sin \psi  } &=& C_1 ( \psi - \psi_* - a \tilde r^2 \sin \psi \cos \psi)\nonu
\tilde x &=& {1 \over \tilde r \sin \psi } - {C_2 \over C_1}\label{solasflat}
\eea
with $C_1,\ C_2$ integration constants that reduce to the previously introduced
 ones (\ref{nearhorsol}) in the $a \to 0$ limit.
The first equation is a quadratic equation for $\tilde r$ and the solutions consist of two branches:
$$
\tilde r_{\pm} = - {1\over 2 a C_1 \cos \psi} \left[ 1 \pm \sqrt{ 1 + 4 a C_1^2 (\psi - \psi_* )\cot \psi}\right]
$$
The two branches join at the point where the argument of the square root becomes zero.
In the near horizon limit $a\to 0 $, only the $-$ branch
survives. The full solution describes a curve starting out at $\psi = \p/2,\ r = \infty$  in the asymptotically flat region
and approaching
the solutions of the previous section  near $r=0$. One should note that, near $r=\infty$, the
radius of the $S^2$ grows like $r^2$ and the solution approaches a flat D3-brane transverse to the D1-D5
string in the background. The transverse distance $\D Y$ between the D3-brane and the D1-D5
string  is given by the limiting value of $r \cos \psi$ as $\psi$ approaches $\p/2$ and
is related to the integration constant $C_1$, which played the role of an asymptotic
modulus in the near-horizon region:
$$\D Y =  \lim_{\psi \to \p/2} | r \cos \psi |=  1/|C_1|.  $$
The general solution can be interpreted as a $(p,q)$ string running between this D3-brane
and the D1-D5 string in the background. This is illustrated  in figures \ref{exampleflow}(b) and \ref{coordflows}(b). \\
\FIGURE{\begin{picture}(300,310)
\put(70,-5){(a)}\put(230,-5){(b)}
\put(20,15){$\tilde r \cos \psi$}\put(180,15){$\tilde r \cos \psi$}
\put(120,20){$\tilde r \sin \psi$}\put(280,20){$\tilde r \sin \psi$}
\put(0,150){$\tilde x$}\put(162,150){$\tilde x$}
\put(70,210){$\tilde r \cos \psi$}\put(220,210){$\tilde r \cos \psi$}
\put(12,220){\epsfig{file=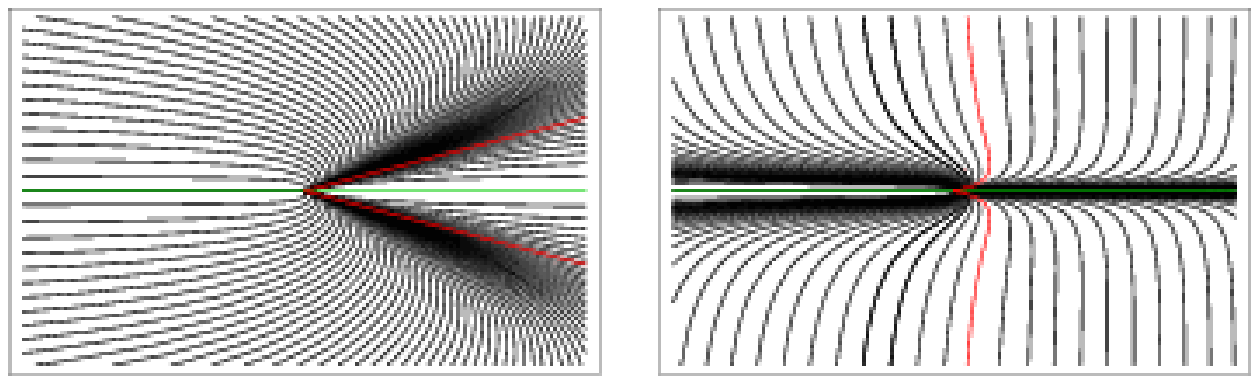, width=300pt }}
\put(-15,260){$\tilde r \sin \psi$}\put(305,260){$\tilde r \sin \psi$}
\put(0,0){\epsfig{file=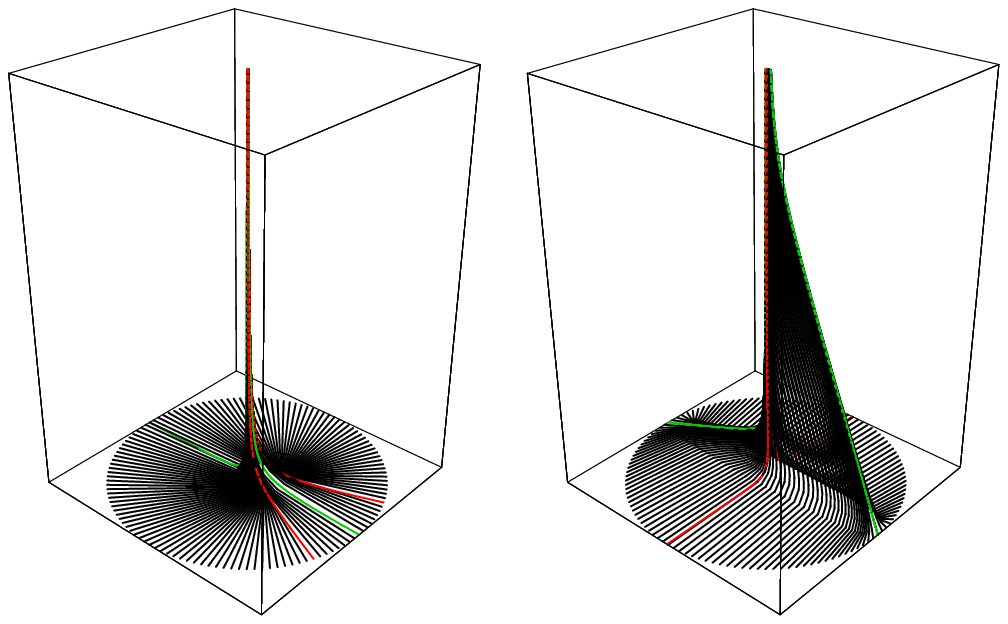, height=200pt }}
\end{picture}\caption{A sampling of  flow solutions plotted in coordinate space (a) in the near-horizon region and
(b) in the full geometry. The upper figure shows the flows projected to the  $(\tilde r \cos \psi, \tilde r \sin \psi )$
plane (as appropriate for the solutions with vanishing D1-charge $p=0$).
The red curve represents the attractor solution, the green
curves are the repulsive fixed point solutions with $\psi=0$ and $\psi = \p$.}\label{coordflows}}
The energy (\ref{energyfull}) of the solutions contains a divergent term as well as a finite one:
$$ E = { 4 \p \m_3 \over g} \lim_{\psi \to \p/2} \left[  {( r(\psi) \sin \psi )^3 \over 3} + r_5^2 (r(\psi) \sin \psi ) \right]
+ ({Q_5 \over 2} - q) \m_1 \D Y .$$
The divergent term is the energy of a flat D3-brane transverse to the D1-D5 string cut off at a large
radius $r_f =r\sin \psi$:
$\m_3 \int e^{-\F} \sqrt{-g} ={4 \p  \m_3\over g} \int_0^{r_f } dr   r^2 (1 + (r_5/r)^2)$. The finite term
represents the energy of ${Q_5 \over 2} - q$ fundamental strings stretched over a distance $\D Y $ perpendicular
to the D1-D5-string.

We can also identify the solutions that reduce to the attractive and repulsive fixed points in the near-horizon
region. These are obtained by taking $C_1 \to \infty$ and $C_1 \to 0$ respectively and correspond to putting the D3-brane
at $r=0$ or $r\to \infty$.
Let's start with the latter case, where the D3-brane has moved off to infinity, leaving behind a  $(p,q)$ string
with the $S^2$ shrunk to zero size. The explicit solution reads
\bea
\psi &=& 0 \nonu
x &=& {r_5^2 p \over g q} {1 \over r} - a {p \over g q} r + C\nonu
F_{t x} &=& {g q \over p H_5}\label{pqasflat}
\eea
It's interesting to look at the large $r$ behaviour:
\bea
x &\sim& -  {p \over g q} r + C\nonu
F_{t x} &\sim& {g q \over p }\nonumber
\eea
This is precisely the solution, in the flat space approximation, of a $(p,q)$ string impinging
on the D1-D5 string in the background, reaching it at an angle $\a$ with $\tan \a = g {q \over p}$
\cite{Dasgupta:1997pu,Rey:1997sp}. This
configuration would arise as the third leg of a three string junction consisting of
$p$ D-strings parallel to the D1-D5 string and  $q$ fundamental strings orthogonal to it, joining up
at $x= C$, as shown in figure \ref{stringjunction}.
\FIGURE{\begin{picture}(300,100)
\put(90,60){(0,q)}\put(120,85){(p,0)}\put(160,0){(p,q)}\put(150,30){$\a$}
\put(90,0){\epsfig{file=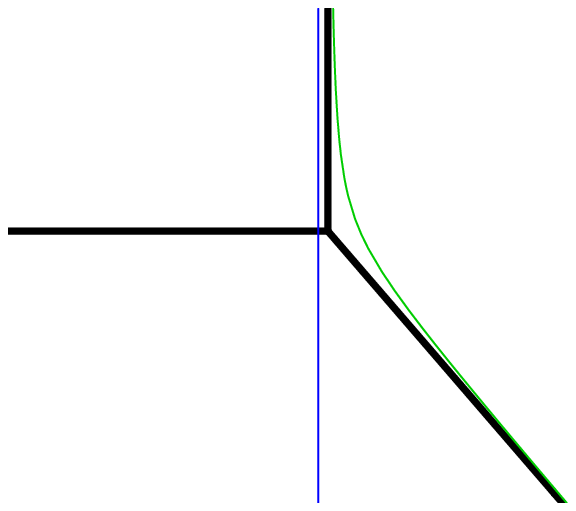, width=150pt }}
\end{picture}\caption{The string junction described in the text. The blue line represents
the D1-D5 string in the background, and the green line shows the bending of the full solution (\ref{pqasflat}).}\label{stringjunction}}
The full solution shows the bending or `kinkiness' \cite{Lambert:1999ix} under the
influence of the
D1-D5 string in the background, moving the junction point off to infinity and leaving only one leg visible.
Hence we can see this solution as the result of placing a $(p,q)$ string junction
in the D1-D5 background. For the general solutions (\ref{solasflat}) a similar interpretation holds,
the only difference being that the $(p,q)$ string leg ends on a D3-brane in these cases.

The solution extending the attractor solution in the near-horizon limit is obtained by moving
the D3-brane to $r=0$. The solution becomes
\bea
\tilde r &=&  \sqrt{ \left| \psi - \psi_* \over \sin\psi \cos \psi \right|}\nonu
\tilde x &=&  \sqrt{ \left|\cot \psi  \over \psi - \psi_*\right|}\nonumber
\eea

\section{Supersymmetry analysis}\label{susyanal}
In this section we show that the BPS equations (\ref{asflatbpseqs}) can alternatively be derived
from the requirement of supersymmetry.
The full supergravity background preserves 8 real `Poincar\'{e}' supercharges, while  in the near horizon
region there are an additional 8 real `conformal' supercharges.
We will show that the BPS D-brane solutions discussed above also display the phenomenon of supersymmetry enhancement:
near $r=0$, supersymmetry is enhanced from 4  Poincar\'{e} supersymmetries preserved by the generic
solution to include an extra 4 conformal supersymmetries preserved by the attractor
solution.

A supersymmetry of the background is preserved in the presence of a
bosonic D-brane configuration if it can be compensated for by a
$\k$-symmetry transformation \cite{Bergshoeff,Bergshoeff:1997bh}. This can be expressed as a projection equation
\be (1 - \G_\k ) \e = 0 \label{kappaproj}\ee where $\G_\k $ (satisfying ${\rm tr} \G_\k  = 0,\ \G_\k ^2 = 1$) is
the operator entering in the $\k$-symmetry
transformation   rule on the D-brane and $\e$ is a general Killing spinor of the background pulled back to the
world-volume.

The Poincar\'{e} Killing spinors of the background can be written as (see appendix \ref{killing} for a derivation in our conventions)
\be
\e = ( H_1 H_5 )^{- 1/8} R(\psi, \theta, \f) \e_0 \label{poincsusies}
\ee
where $R(\psi, \theta, \f)$ is a rotation
$$ R (\psi, \theta, \f) = e^{ { {\p \over 2 } - \psi \over 2} \G^{\utheta \uphi} \s^1} e^{ { {\p \over 2 } - \theta \over 2}
 \G^{\upsi \uphi} \s^1} e^{ {\f \over 2} \G^{\upsi \utheta} \s^1}.$$
The spinor $\e_0$ is constant on the six-dimensional space, covariantly constant on $\cm$ and satisfies the projection
equations
\bea
\left( 1 + \G^{\ut\ux} \s^1\right)\e_0 &=& 0 \nonu
\left( 1 - \G^{\ur \upsi \utheta \uphi }\s^1 \right) \e_0 &=& 0 \label{poincproj}
\eea

The operator $\G_\k $ entering in (\ref{kappaproj}) depends on the embedding of the D3-brane in the background as well
as on the worldvolume gauge field. As before, we take the worldvolume coordinates to be $(t,x,\theta,\f)$.
Imposing  $SU(2)$ symmetry as discussed in section \ref{bpsder}, the induced worldvolume metric on a static D3-brane is
\bea
d\hat s^2 &=& - (H_1 H_5)^{-1/2} dt^2 + \left[ (H_1 H_5)^{-1/2} + (H_1 H_5)^{1/2} (r'^2 + r^2 \psi'^2) \right]
dx^2 \nonu
&&+ r^2 (H_1 H_5)^{1/2} \sin^2 \psi ( d \theta^2 + \sin^2 \theta d\f^2 ) .\nonumber
\eea
where a prime denotes a derivative with respect to $x$.
The form of the  worldvolume gauge fields is restricted by fixing the $(p,q)$ charge and was given in
the expressions (\ref{magnfield}, \ref{elfield}). In terms of the vielbein for the above induced metric the
gauge fields read
$$ F = {\D_1 \over \sqrt{\D_1^2 + \D_2^2 + \D_3^2 }} e^{\hut}\wedge e^\hux + {\D_3 \over \D_2} e^\hutheta \wedge
e^\huphi$$
We use an index convention where a hatted index denotes a pullback to the worldvolume and orthonormal frame
indices are underlined.
The $\k$-operator $\G_\k $ is then \cite{Bergshoeff,Bergshoeff:1997bh}
$$ \G_\k = e^ {- \F_0 \G^{\hut\hux} \s^3 - \F_1 \G^{\hutheta\huphi} \s^3} \G_{\hut\hux\hutheta\huphi} i \s^2$$
with $\F_0,\ \F_1$ defined by
$$ \tanh \F_0 = {\D_1 \over \sqrt{\D_1^2 + \D_2^2 + \D_3^2 }};\qquad \tan \F_1 = {\D_3 \over \D_2} .$$
The pulled-back gamma matrices are related to the 10-dimensional ones as
\bea
\G_{\hut\hux}&=&{1 \over \sqrt{1 + (H_1 H_5) (r'^2 + r^2 \psi'^2)}}\left(  \G_{\ut\ux} +
 (H_1 H_5)^{1/2} r' \G_{\ut\ur} +   (H_1 H_5)^{1/2} r \psi' \G_{\ut \upsi} \right)\nonu
 \G_{\hutheta\huphi}&=&\G_{\utheta\uphi}\nonumber
\eea
Requiring $\G_\k  \e = \e$ with $\e$ given in (\ref{poincsusies}) for all values of $\theta$ and $\f$ leads to
two equations which can be summarized as
\be \left( 1 -  e^{s \F_0 \G_s \s^3} e^{-\F_1 \G^{\utheta\uphi}\s^3} \G_s \G^{\utheta\uphi}i \s^2 \right)
e^{ s{ {\p \over 2 } - \psi \over 2} \G^{\utheta \uphi}\s^1} \e_0 = 0\label{kappaeqs}\ee
where  we defined the operator $\G_s$
$$\G_s = {1 \over \sqrt{1 + H_1 H_5 (r'^2 + r^2 \psi'^2)}}\left(  \G_{\ut\ux} +
 (H_1 H_5)^{1/2} r' \G_{\ut\ur} + s  (H_1 H_5)^{1/2} r \psi' \G_{\ut \upsi} \right) $$
and $s$ can be $1$ or $-1$.
Some algebraic manipulations reduce the equations (\ref{kappaeqs}) to the following system
\bea
(H_1 H_5)^{1/2} \D_3 r\psi' &=& \pm [\cos \psi \D_2 - \sin \psi \D_1 ]\nonu
(H_1 H_5)^{1/2} \D_3 r ' &=& \pm [\sin \psi \D_2 + \cos \psi \D_1 ]\nonu
(1 \pm \G_{\ut\upsi} \s^3 ) \e_0 &=&0 \label{susyeqs}
\eea
The first two equations are identical to (\ref{asflatbpseqs}), as one can see by making use of
\bea
\pa_r (r Z ) &=& \sin \psi \D_2 + \cos \psi \D_1 \nonu
\pa_\psi  Z  &=& \cos \psi \D_2 - \sin \psi \D_1 \nonumber
\eea
The projector in (\ref{susyeqs})  commutes with (\ref{poincproj}), showing that the solutions preserve 4 out of the 8
Poincar\'{e} supersymmetries. Note that the solutions with different sign choices in (\ref{susyeqs}) are not mutually BPS.

We now proceed to verify whether, in the near horizon limit $a=0$, any of the solutions preserve
some of the enhanced conformal supersymmetries. These are given by (see appendix \ref{killing} for details):
\be
\tilde \e = \left( {1\over\sqrt{u}} +  \sqrt{u}( t \G^{\ut\uu} - x \G^{\ux\uu})\right)
  R(\psi, \theta, \f)\tilde \e_0 \label{enhancedsusies}
\ee
where we defined a rescaled radial coordinate $u \equiv {r\over r_1 r_5}$.
The spinor $\tilde \e_0$ is constant on $AdS_3 \times S^3$, covariantly constant on $\cm$ and satisfies the projection
equations
\bea
\left( 1 - \G^{\ut\ux} \s^1\right)\tilde \e_0 &=& 0 \nonu
\left( 1 + \G^{\ur \upsi \utheta \uphi }\s^1 \right)\tilde \e_0 &=& 0 .\label{enhancedproj}
\eea
We look for solutions of $(1 - \G_\k ) \tilde \e=0$, where $\G_\k$ is as in (\ref{kappaproj}) with $a$ put to zero in the
harmonic functions. In particular, the pulled-back gamma matrices are
\bea
\G_{\hut\hux}&=&{1 \over \sqrt{1 + {1\over u^4} (u'^2 + u^2 \psi'^2)}}\left(  \G_{\ut\ux} +
 {u'\over u^2}  \G_{\ut\uu} +   {1\over u}  \psi' \G_{\ut \upsi} \right)\nonu
 \G_{\hutheta\huphi}&=&\G_{\utheta\uphi}\nonumber
\eea
Since the background Killing spinor $\tilde \e$ is time dependent and we are looking for static solutions
of $(1 - \G_\k) \tilde \e=0$, the only possibility is for the coefficient of $t$ in this equation to vanish separately.
This leads to two equations
\bea
(1 - \G_\k) \G^{\ut\uu} R(\psi, \theta, \f)\tilde \e_0 &=&0 \\
(1 - \G_\k) (\G^{\ut\uu} - u x \G^{\ut\ux}) \G^{\ut\uu} R(\psi, \theta, \f)\tilde \e_0 &=&0
\eea
Making use of the properties of $\tilde \e_0$ in (\ref{enhancedproj}), one can show that the first equation
leads  to the equations we had before in (\ref{susyeqs}), with $\e_0$ replaced by $\tilde \e_0$. If these
are satisfied, the second equation becomes equivalent to
\be [\G_{\hut\hux},(\G_{\ut\uu} - u x \G_{\ut\ux})] \G^{\ut\uu} R(\psi, \theta, \f)\tilde \e_0  =0.\label{enhancedcond}\ee
The commutator equals
$$ [\G_{\hut\hux},(\G_{\ut\uu} - u x \G_{\ut\ux})] = {2\over \sqrt{1 + {1\over u^4} (u'^2 + u^2 \psi'^2)}}\left(
(1 + {xu'\over u})\G_{\ux\uu} + {\psi'\over u}\G_{\upsi\uu} - x \psi' \G_{\upsi\ux}\right) .$$
The equation (\ref{enhancedcond}) requires $\det  [\G_{\hut\hux},(\G_{\ut\uu} - u x \G_{\ut\ux})]^2 =0 $
which leads to
$$ (1 + {xu'\over u})^2 + \left({\psi'\over u}\right)^2 + (x\psi')^2 = 0 .$$
Hence we see that the solutions preserving conformal supersymmetries have to satisfy
\bea
1 +{xu'\over u}&=& 0 \nonu
\psi' &=& 0. \nonumber
\eea
This singles out the attractor solution found in (\ref{nearhorattrsol}). It preserves 4 extra conformal supersymmetries
specified by  the projection
$$ (1 \pm \G_{\ut\upsi} \s^3 ) \tilde \e_0 =0. $$
The supersymmetry preservation of the attractor solution was studied before in \cite{Raeymaekers:2006np}.

\section{S-dual solutions}\label{sdualtransf}

The  solutions of in sections \ref{studynear}, \ref{asflatspace}  can of course be transformed to
solutions in different duality frames for the background 2-charge
system. Of special interest is the S-dual F1-NS5 background composed of fundamental strings and Neveu-Schwarz
fivebranes. This background can, in the near-horizon region, be described as an exact conformal field theory
on the $SL(2,R) \times SU(2)$ Wess-Zumino-Witten model at level $Q_5$. We will now describe how our solutions
transform under S-duality.

The background geometry transforms into
\bea
e^{-\F'} &=& e^{\F} = {1 \over g'}\left( { H_1 \over H_5 } \right)^{1/2}; \qquad g' = 1/g\nonu
ds'^2 &=& e^{-\F} ds^2= g' \left[ (H_1)^{-1} ( - dt^2 + dx^2 ) + H_5 (dr^2 + r^2 d\O_3^2) +
\left( { H_1 \over H_5 } \right) ds^2_\cm  \right] \nonu
B'^{(2)} &=& C^{(2)} = {g' \over  H_1 } dt \wedge dx + {g' r_5^2}
( \psi - \sin \psi \cos \psi ) \sin \theta d\theta \wedge d\f \label{Sdualbackground}
\eea

The transformed D3-brane solutions have an induced metric given by $d\hat s'^2 = e^{-\F} d\hat s^2$, while the
equations of motion and the Bianchi identities for the worldvolume gauge field are reversed
\cite{Gibbons:1995ap,Tseytlin:1996it,Green:1996qg}:
\bea
F' &=& \star K \nonu
\star K' &=& - F \label{Ftransf}
\eea
where $K$ is defined as
$$ K^{\m\n} \equiv {1 \over \m_3 \sqrt{-\det \hat G}} {\d S \over \d F_{\m\n}}. $$
The charge quantization conditions (\ref{f1quant}, \ref{d1quant}) then imply
\bea
q  &=& {\m_3 \over \m_1} \int_{S^2} F'\nonu
p  &=& - {\m_3 \over \m_1} \int_{S^2} \star K'. \nonumber
\eea
These are the  usual quantization conditions  for a $(D1,F1)=(q,p)$ string in the F1-NS5 background
\cite{Bachas:2000ik,Taylor:2000za}.
The fact that, in the near-horizon limit,  $q$ is a ${\bf Z}_{Q_5}$-valued charge is also well-established in this case
\cite{Alekseev:2000jx,Fredenhagen:2000ei,Maldacena:2001xj,Maldacena:2001ss}.
Applying the transformation (\ref{Ftransf}) one finds
\bea
F'_{\theta \f} &=& - {q \m_1 \over 4 \p \m_3} \sin \theta \nonu
F'_{t x} &=& 0\nonumber
\eea
Defining
$ L'^2 = Q_5 \a'$, the induced metric  and the gauge-invariant field strength $\cf = F + B$ become
\bea
d \hat s^2 &=& {g' \over H_1}  \left( - dt ^2 + {  \D_1^2 + \D_2^2  +\D_3^2 \over \D_3^2} d x^2 \right)
+L'^2 \D_2 \left( d \theta ^2 + \sin^2 \theta d \f^2\right)\nonu
\cf &=& {g' \over H_1}  dt \wedge d x - L'^2 \D_1 \sin \theta d \theta \wedge d\f.
\eea
For the open string metric $g^o_{\m\n} = g_{\m\n} - \cf_{\m\r} g^{\r\s}\cf_{\s\n}$, one finds:
\be
d s_o^2 = {g'(\D_1^2 + \D_2^2)\over H_1} \left( - {dt^2 \over  \D_1^2 + \D_2^2  +\D_3^2}
+  { dx^2  \over \D_3^2}\right)  +{L'^2 (\D_1^2 + \D_2^2)\over \D_2}\left( d\theta^2 + \sin^2 \theta d\f^2 \right)
\label{Sdualopenmetric}
 \ee

To evaluate these formulas, one has to use the definitions (\ref{defdeltas}) together with the solutions (\ref{nearhorsol})
or (\ref{solasflat}). For example, one easily checks that, in the near-horizon limit, the attractor solution is $AdS_2 \times
S^2$ with radii in the induced metric given by given by
\be R_{\rm AdS_2} = L' {{T'}_{(q,p)}^{\rm attr} \over {T'}_{(q,0)}^{\rm attr}} ;\qquad R_{\rm S^2}
= L' Z_* \ee
where we defined
$$ {T'}_{(q,p)}^{\rm attr} \equiv \m_1 \sqrt{p^2 + ({Q_5 \over \p} \sin  {p \p \over Q_5})^2 e^{- 2 \F'}}. $$
In the open string metric (\ref{openmetric}), the radii become equal to the radius of the background geometry:
\be
R_{\rm AdS_2}^o = R_{\rm S_2}^o = L'.
\ee
This particular solution was first studied in  \cite{Bachas} and the above agrees  with the results obtained there.

\section{Discussion}\label{disc}

In this paper we have discussed a simple example of a supersymmetric attractor mechanism in an
open string setting. We found many  similarities with the well-known
attractor mechanism for supersymmetric black holes.
This  raises quite a few questions to which we have not given a satisfactory answer.

An obvious (but perhaps naive) question is  whether, like in the case of  black holes, the physics underlying the
attractor mechanism is related to a microscopic entropy carried by open string degrees of freedom.
A  better understanding of the connection between the open string attractor
mechanism and the open string version of the OSV conjecture, proposed in \cite{Aganagic:2005dh} in a different setting,
 could  shed  light on this issue.

Another question concerns the generalization of the  mechanism to other backgrounds
and  the identification of the general conditions under which it occurs. As in the closed string case,
one would expect a close relation to the mechanism for open string moduli stabilization which was
discussed in \cite{Gomis:2005wc}. As remarked in the Introduction, one would also not expect the mechanism to be
restricted to supersymmetric cases.

For black holes, special geometry plays an important role in the attractor mechanism.
It would  be useful to gain more insight in the supersymmetric geometry underlying the attractor mechanism
in the present example. This would require a better understanding of the how the superconformal symmetry algebra
of the background gets realized nonlinearly on the D3-brane worldvolume.
 One of the quantities for which we would like to have a better interpretation is the function which we
called $Z$ and which controls the attractor flow.

After  S-dualizing, we found  brane solutions in the near-horizon limit of the
F1-NS5 background. These should be describable as boundary states which cannot be obtained
by tensoring together $SL(2,R)$ and $SU(2)$ boundary states, and it would be interesting to obtain these.

And finally, our attractor flow solutions in the near-horizon limit of the D1-D5 system merit an interpretation from
the point of the dual CFT. All solutions run to the boundary of $AdS_3$, which they intersect in a line.
Hence they should correspond to line defects in the CFT, generalizing the ones studied in \cite{Bachas:2001vj}.
Similar branes in the $AdS_5 \times S^5$ background were given a dual CFT interpretation as Wilson lines
in an antisymmetric representation in \cite{Gomis:2006sb,Gomis:2006im}. It would also be of interest to construct the
`bubbling' solutions incorporating backreaction.

\acknowledgments{It is a pleasure to thank K.P. Yogendran, Y. Matsuo, G. Mandal,
Y. Sugawara, Y. Imamura, S. Minwalla, S. Trivedi, A. Ramallo  and especially F. Denef
for useful suggestions and discussions.
I would also like to thank the Tata Institute of Fundamental Research, where part of this work was completed, for
hospitality.}

\appendix

\section{Description as fuzzy $(p,q)$ strings}\label{myers}
Our D3-brane configurations have, in a certain regime of the parameters, an equivalent description as $(p,q)$  strings expanded
to form a D3 brane on a fuzzy $S^2$. We will now describe this version of the Myers effect and show that, in the relevant
regime, the solutions  arise from the noncommutative worldvolume theory for $p$ D-strings in the D1-D5 background.

We start by introducing auxiliary Cartesian coordinates $Y^i,\ i=1,2,3$ satisfying the constraint $\sum_i (Y^i)^2 = 1$
such that the volume element on the $S^2$ can be written as
$$\sin \theta d \theta \wedge d \f = \half \e_{ijk} Y^i d Y^j \wedge d Y^k.$$
We now consider Myers' action \cite{Myers:1999ps} for $p$ D-strings in the given background. We choose a static gauge
where the worldvolume is parametrized by $t,x$. The worldvolume fields now become $p\times p$ matrices.
We will restrict attention to static configurations, starting from an ansatz  of the form
\bea
{  F}_{tx} = F_{tx}(x) { \bf 1}_{p\times p}\nonu
{  r} =r(x) {\bf  1}_{p\times p}\nonu
{  \Psi} =\psi(x) { \bf 1}_{p\times p}.\nonumber
\eea
Furthermore, we take ${  Y}^i$ to be arbitrary constant matrices satisfying the constraint $\sum_i ({  Y}^i)^2 = 1$.
The latter can be implemented by introducing a (matrix-valued) Lagrange multiplier $\l$. The multi-D1 brane action at leading
order then takes the form \cite{Myers:1999ps}
\bea
S&=& - \m_1 {\rm Tr} \int dt dx \Big[ e^{-\F} \sqrt{ - \det (P[G_{ab}] + {  F}_{ab})} \sqrt{\det {  Q}^i_j} \nonu
&& + C^{(2)}_{tx} + {i \over 2 \p\a '}i_Yi_Y C^{(2)} {  F}_{tx} + \l (\sum_i ({  Y}^i)^2 - 1)\Big] \label{noncommaction}
\eea
where
\bea
 \sqrt{ - \det (P[G_{ab}] + {  F}_{ab})} &=& \sqrt{ (H_1 H_5)^{-1} + r'^2 + r^2 \psi'^2 - F_{tx}^2}\nonu
\sqrt{\det {  Q}^i_j} &=& 1 - {H_1 H_5 r^4 \sin^4 \psi \over 4 (2 \p \a')^2 }\sum_{i,j} [{  Y}^i,{  Y}^j ]^2
\nonu
i_Yi_Y C^{(2)} &=& { r_5^2 \over 2 g} ( \psi - \sin \psi \cos \psi ) \e_{ijk} {  Y}^i [ {  Y}^j , {  Y}^k ] \nonumber
\eea
The equations of motion for the matrices ${  Y}$ following from this action are
\bea
0&=&  {1 \over g}\sqrt{ H_5 \over H_1 }\sqrt{ (H_1 H_5)^{-1} + r'^2 + r^2 \psi'^2 - F_{tx}^2}
 {H_1 H_5 r^4 \sin^4 \psi \over  (2 \p \a')^2 } \sum_{j}{  Y}^j  [{  Y}^i,{  Y}^j ]\nonu
&& + {3 i  r_5^2 \over 4\p \a '  g}
( \psi - \sin \psi \cos \psi ) \e_{ijk}[ {  Y}^j , {  Y}^k ]F_{tx} + 2 \l {  Y}^i \label{Yeqs}
\eea
When the ${  Y}$'s are taken to form a fuzzy two-sphere, $[ {  Y}^i , {  Y}^j ] \sim i \e_{ijk} {  Y}^k$,
one sees that each term in this equation is proportional to ${  Y}^i$. Hence the equation is trivially solved
by adjusting  the Lagrange multiplier. In other words, the variation of the action around a fuzzy sphere configuration
is proportional to $\sum Y^i \d Y^i$ and vanishes for variations on the constraint surface. The necessary ingredients that
went into this are the fact that the
auxiliary $Y^i$-space is flat, and that the background magnetic potential $C^{(2)}_{\rm magn}$ is constant over the $S^2$.

In terms of matrices ${  t}^i$ in the $p$-dimensional irreducible representation of $SU(2)$, satisfying
$[ {  t}^i , {  t}^j ] = - i \e_{ijk} {  t}^k$, the ${  Y}$'s are
$$ {  Y}^i = {2 \over \sqrt{ p^2 -1}} {  t}^i.$$
Substituting into the action (\ref{noncommaction}) leads to
\bea
S &=& - p \m_1  \int dt dx  \Big[  {1 \over g}\sqrt{ H_5 \over H_1 }\sqrt{ (H_1 H_5)^{-1} + r'^2 + r^2 \psi'^2 - F_{tx}^2}
\left( 1 +  {2 H_1 H_5 r^4 \sin^4 \psi \over (p^2 -1) (2 \p \a')^2 }\right) \nonu
&& + {r_1^2 \over g H^1} + {Q_5 \over \sqrt{p^2 -1} \p} ( \psi - \sin \psi \cos \psi )F_{tx} \Big].\nonumber
\eea
We see that this expression agrees with (\ref{action}) for static configurations  in the large $p$ limit when
$$
{H_1 H_5 r^4 \sin^4 \psi \over p^2 (\a' \p)^2 }\ll 1.
$$

The equations (\ref{Yeqs}) also allow for a solution where the the $Y^i$ are commuting, forcing the Lagrange
multiplier $\l$ to be zero. This represents a $(p,q)$ string which has not expanded into a fuzzy 2-sphere.
The existence of solutions for both expanded and non-expanded configurations is similar to the case of
giant gravitons \cite{McGreevy:2000cw,Grisaru:2000zn}.

\section{Killing spinors}\label{killing}
In this appendix we derive the form of the Killing spinors in the D1-D5 background.
We follow the conventions of \cite{Bena}, in which the type IIB gravitino and dilatino variations (for
vanishing $H,\ F^{(1)}, F^{(5)}$) are given by
\bea
\d \l &=& \left[\half \G^M\pa_M - {1 \over 4} e^\F  \sF{}^{(3)} \s^1 \right]\e \nonu
\d \Psi_M &=& \left[ \nabla_M + {1\over 8} e^\F  \sF{}^{(3)}\G_M \s^1 \right]\e .\nonumber
\eea
Here, $\e$ is a doublet of chiral spinors in 10 dimensions with chirality $\G_{(11)} \e \equiv \G^{\underline{0}\ldots
\underline{9}}\e= -\e$.

The vanishing of the dilatino variation in the background (\ref{background})   imposes the following projection equations:
\bea
\left( 1 + \G^{\ut\ux} \s^1\right)\e &=& 0 \nonu
\left( 1 - \G^{\ur \upsi \utheta \uphi }\s^1 \right) \e &=& 0 \nonumber
\eea
where we use underlined indices to denote  orthonormal frame indices.
This projects the number of independent real components of $\e$ down to 8.
The near-horizon background $a=0$ allows extra solutions which give rise to 8 enhanced supersymmetries and which
will be discussed below.

The vanishing of the gravitino variation determines the coordinate dependence of $\e$. The components on
the internal manifold $\cm$ lead to the condition the $\e$ is covariantly constant with respect to the Ricci-flat metric
on $\cm$. The 6-dimensional components have the solution
\be
\e = ( H_1 H_5 )^{- 1/8} R(\psi, \theta, \f) \e_0
\ee
where $R(\psi, \theta, \f)$ is a rotation
$$ R (\psi, \theta, \f) = e^{ { {\p \over 2 } - \psi \over 2} \G^{\utheta \uphi} \s^1}
e^{ { {\p \over 2 } - \theta \over 2} \G^{\upsi \uphi} \s^1} e^{ {\f \over 2} \G^{\upsi \utheta} \s^1}.$$
The spinor $\e_0$ is constant in 6 dimensions, covariantly constant on $\cm$ and satisfies the projection
equations $\G^{\ut\ux} \s^1 \e_0 = - \e_0,\  \G^{\ur \upsi \utheta \uphi }\s^1 \e_0 = \e_0$.

In the near-horizon limit $a=0$, the dilatino equation allows extra solutions $\tilde \e$ satisfying
\bea
\left( 1 - \G^{\ut\ux} \s^1\right)\tilde \e &=& 0 \nonu
\left( 1 + \G^{\ur \upsi \utheta \uphi }\s^1 \right) \tilde \e &=& 0 \nonumber
\eea
The gravitino equations then lead to the following form of these enhanced supersymmetries:
\be
\tilde \e = \left( \sqrt{ r_1 r_5 \over r} +  \sqrt{ r \over r_1 r_5}( t \G^{\ut\ur} - x \G^{\ux\ur})\right)
  R(\psi, \theta, \f)\tilde \e_0
\ee
where $\tilde \e_0$ is constant on $AdS_3 \times S^3$, covariantly constant on $\cm$ and satisfies the projection
equations $\G^{\ut\ux} \s^1 \tilde \e_0 = \tilde \e_0,\  \G^{\ur \upsi \utheta \phi }\s^1 \tilde \e_0 =- \tilde \e_0$.

\end{document}